\documentclass[a4paper,12pt]{article}
\usepackage{exscale}
\usepackage{amssymb,amsmath,latexsym,amsthm,amscd}

\renewcommand{\theequation}{\arabic{section}.\arabic{equation}}

\begin{document}

\title{\bf osp(1,2)--covariant Lagrangian quantization of irreducible 
massive gauge theories with generic background configurations}
\vspace{1cm}
\author{\large{Bodo~Geyer}\thanks{e-mail: geyer@itp.physik.uni-leipzig.de}
\\
\emph{Universit\"at~Leipzig, Institut~f\"ur~Theoretische~Physik}
\\
\emph{Augustusplatz~10, D-04109~Leipzig, Germany}
\\[2ex]
{Dietmar M\"ulsch} 
\\
\emph{Wissenschaftszentrum Leipzig e.V., }\\
\emph{Goldschmidtstr.~26, D--04103 Leipzig, Germany}}
\date{}

\maketitle
\thispagestyle{empty}

\begin{abstract}
\noindent
In the framework of the $osp(1,2)$-symmetric 
quantization of irreducible massive gauge theories the background 
field method is studied for the simplest case of a linear splitting of
the gauge field into a background configuration $A^i$ and the quantum
fluctuations $Q^i$. The entire set of symmetries of that approach,
including three types of background--dependent gauge transformations,
is expressed by the corresponding set of Ward identities. Making use 
of these identities, together with the equation of motion of the
auxiliary field, the background dependence of the generating functionals of
the 1PI vertex functions and the Green's functions is completely determined 
by the dependence of these functionals 
upon the gauge fields and the associated antifields. 
It is proven that the introduction of a background field  does not 
change the ultraviolet asymptotics of the theory.
\end{abstract}


\section{Introduction and general results}
\renewcommand{\theequation}{\thesection.\arabic{equation}}
\setcounter{equation}{0} 


Lagrangian quantization of general gauge theories is usually considered 
in the Batalin--Vilkovisky(BV) approach \cite{1,2} as well as 
in the $Sp(2)$--covariant formalism of Batalin, Lavrov and Tyutin \cite{3,4} 
which is a generalization of the former observing the extended 
BRST--symmetry, the symmetry under BRST-- and antiBRST--transformations. 
Recently, the $Sp(2)$--covariant quantization has been extended to a formalism
which is based on the orthosymplectic superalgebra $osp(1,2)$ \cite{5} and 
which can be applied to {\it massive} theories. 

The very general quantization procedures \cite{1} -- \cite{4} may be applied 
also to gauge theories with external 
fields and/or composite fields \cite{6}. However, there is a wide class of 
physically interesting situations where these procedures seem not to be 
optimally adapted. This is the case if quantum effects of gauge fields in 
the presence of classical gauge configurations, like instantons or merons or 
monopoles, have to be evaluated or, more simply, if the background field 
method (BFM) \cite{7} should be applied in order to facilitate explicit 
computations. 

The BFM is very convenient also for the investigation of general properties 
of gauge theories and, therefore, deserves a comprehensive study.
Among the many attractive features of the BFM one should mention that by 
chosing a background covariant gauge it is possible to compute quantum 
effects without losing manifest gauge invariance. This is due to the fact 
that in its original formulation the BFM introduces a {\em linear splitting} 
of the gauge field into quantum fluctuations $Q$ around the background 
field $A$ such that the gauge transformation of the latter may be chosen 
inhomogeneous whereas the gauge transformations of the quantum field occure 
homogeneous like those of the (anti)ghost and the matter fields. 

The BFM may also be used in the proof of renormalizability of gauge 
theories within the BRST approach. Concerning the role of the background
field an essential step has been made by Kluberg-Stern and Zuber \cite{8} 
distinguishing between BRST and type--I transformations. In principle, 
the procedure of quantizing gauge fields in the background of a generic
classical field configuration, as has been used in Ref.~\cite{8}, could be  
taken over to the BV--formalism by chosing background covariant gauges. 
However, since the BV--formalism essentially works in the 
{\it minimal sector} of the theory -- the nonminimal sector being trivial --  
one will not be able to make use of the full symmetry content of the 
classical theory.  

Therefore, the aim of the present paper is to offer some general aspects of 
the BFM in the $osp(1,2)$--covariant quantization scheme where the results 
of Ref.~\cite{8} are extended nontrivially to the 
non--minimal sector of the theory (A short exposition of that approach 
already has been given in Ref.~\cite{9}; in a more restricted context these
problems have been considered in Ref.~\cite{10}). In that approach
all symmetries of the classical action can be imposed as Ward identities
also for the full quantum action. This leads, on the one hand, to the
symmetries of the $osp(1,2)$--superalgebra and, on the other hand, to 
the so-called type I -- type III symmetries related to the background field.

Here, we prefer the $osp(1,2)$--covariant quantization scheme, 
rather than the $Sp(2)$--covariant formalism, since the quantization of 
gauge theories in the presence of classical background configurations, like
instanton or meron solutions, may be plagued by severe infrared problems
which requires the consideration of massive gauge fields. On the other hand,
the incorparation of mass terms into the quantum action is necessary, 
at least intermediately, in the renormalization scheme of Bogoliubov, 
Parasiuk, Hepp, Zimmermann and Lowenstein(BPHZL) \cite{11}. In order to 
deal with massless theories in that scheme an essential ingredient 
consists in the 
introduction of a regularizing mass $m_s = (s - 1) m$ for any massless field 
and performing ultraviolet as well as infrared subtractions thereby avoiding 
spurious infrared singularities in the limit $s \rightarrow 1$. By using that 
infrared regularization -- without violating the extended BRST symmetries -- 
the $osp(1,2)$--superalgebra occurs necessarily. Moreover, the BPHZL
renormalization scheme is surely the mathematical best founded one in 
order to formulate the quantum master equations on the level of algebraic
renormalization theory and to properly compute higher--loop anomalies 
\cite{12}.

Now, to make our assertions more explicit let us relate our method to 
the realm of general gauge theories, thereby introducing also the necessary
prerequisites of gauge theories with a generic background field.
In order to simplify the complexity of the $osp(1,2)$--formalism, we 
restrict ourselfes to the consideration of irreducible gauge theories of 
first rank with semi--simple Lie group in 4--dimensional space--time. 
Thorough this paper we use the condensed notation 
introduced by DeWitt \cite{13} and conventions concerning the derivatives 
with respect to fields and antifields adopted in Ref.~\cite{3}. However, 
for later notational simplicity all the {\em antifields} are introduced 
with {\em opposite sign} in comparison with Ref.~\cite{1} -- \cite{5}!

We consider a set of gauge fields $G^i$ which, for the sake of
simplicity, are supposed to be {\em bosonic} and whose classical action
$S_{\rm cl}(G)$ is invariant under the gauge transformations
\begin{equation}
\label{1.1}
\delta G^i = R^i_\alpha(G) \xi^\alpha,
\qquad
R^i_\alpha(G) S_{{\rm cl}, i}(G) = 0
\qquad
\hbox{with}
\qquad
S_{{\rm cl}, i}(G) \equiv \frac{\delta}{\delta G^i} S_{\rm cl}(G),
\end{equation}
where $\xi^\alpha$ and $R^i_\alpha(G)$ are the parameters of the gauge
transformations and the gauge generators, respectively. 
We assume the set of generators $R^i_\alpha(G)$ to be linearly independent
and complete (irreducible or zero--stage theories). Furthermore, to be able to 
introduce the background field $A^i$ by a {\it linear quantum--background 
splitting} of the gauge fields, $G^i = A^i + Q^i$, we make the assumption that 
$R^i_\alpha(G)$ depends on $G^i$ only {\it linearly},
\begin{equation}
\label{1.2}
R^i_\alpha(G) = R^i_\alpha(0) + R^i_{\alpha, j} G^j,
\qquad
R^i_\alpha(0) \equiv r^{i \mu}_\alpha \partial_\mu,
\end{equation}
and, for the sake of renormalizability, that $R^i_\alpha(0)$ are {\it linear}
differential operators.
(Notice, that supersymmetric theories in general do not obey this 
restriction but that the majority of physically interesting theories, 
including, e.g., $N = 2$ super Yang--Mills theories in harmonic superspace, 
comply it.) The (closed) gauge algebra is expressed by 
\begin{equation}
\label{1.3}
R_{\alpha, j}^i R_\beta^j(G) - R_{\beta, j}^i R_\alpha^j(G) =
- R_\gamma^i(G) {f^\gamma}_{\alpha\beta}.
\end{equation}
Here, $f_{\alpha\beta}^\gamma$ are the structure constants of a semi-simple 
Lie group which obey the usual Jacobi identity and, by the help of the 
(nonsingular) Killing metric 
$g_{\alpha\beta} \equiv f_{\alpha\gamma}^\delta f_{\beta\delta}^\gamma$,
may be chosen totally antisymmetric after lowering all the indices.
Furthermore, introducing the metric tensor with respect to the field 
components, $g_{ij}$, with $g_{ij} g^{jk} = \delta^k_i$, the following 
similar relations hold:
$g_{ki} R^i_{\alpha, j} = - g_{ji} R^i_{\alpha, k}$,
$g^{ik} R^j_{\alpha, i} = - g^{ij} R^k_{\alpha, i}$;
of course, they are compatible with the gauge algebra (\ref{1.3})
(observe that according to DeWitt's notation $i = (\mu, \alpha, x, \ldots)$
summarizes Lorentz and group indices, space-time points, etc.).  

Now we point to the fact that, because of the linear quantum--background
splitting the symmetry transformations (\ref{1.1}) of $S_{\rm cl}(A + Q)$ 
can be realized in three different ways: as {\it background} or type--I gauge 
transformations \cite{11}, and two forms of gauge transformations of the 
{\it quantum field}, one with $A^i$ held fixed (called type--II) and 
another one with $A^i$ transforming homogenously (called type--III):
\begin{alignat}{2}
\label{1.4}
\hbox{type--I}: 
\qquad
\delta A^i &= R^i_\alpha(A) \xi^\alpha,
&\qquad
\delta Q^i &= R^i_{\alpha, j} Q^j \xi^\alpha,
\\
\label{1.5}
\hbox{type--II}:
\qquad 
\delta A^i &= 0,
&\qquad
\delta Q^i &= R^i_\alpha(A + Q) \xi^\alpha,
\\
\label{1.6}
\hbox{type--III}:
\qquad
\delta A^i &= R^i_{\alpha, j} A^j \xi^\alpha,
&\qquad
\delta Q^i &= R^i_\alpha(Q) \xi^\alpha.
\end{alignat}
Notice that all these symmetries are written down with the parameters
$\xi^\alpha$ of the gauge transformations. To our knowledge, so far
neither type--II nor type--III symmetry have been considered in the 
literature as {\em quantum symmetries}. However, this is possible if 
the theory, as in our case, is appropriately extended to the 
{\it non--minimal sector!}
Thereby, in order to remove the gauge degeneracy of the classical 
action $S_{\rm cl}(A + Q)$, one should choose the gauge fixing term in such 
a way that the invariance under type--II and type--III transformations is 
fixed whereas its invariance under type--I transformations, 
i.e.,~under a gauge change of the classical background, is not to be affected. 

In the $Sp(2)$--covariant Lagrangian quantization \cite{3} the effective
(gauge fixed) action $S_{\rm eff}(A| \phi)$ with 
$\phi^I = ( Q^i, B^\alpha, C^{\alpha b} )$ denoting the (quantum part of the)
gauge field, the auxiliary field and the (anti)ghost fields, respectively,
is invariant under properly generalized type--I and 
(anti)BRST--transformations. Introducing for each field $\phi^I$ two 
additional sets of antifields 
$\phi_{I a}^* = ( Q_{i a}^*, B_{\alpha a}^*, C_{\alpha ab}^* )$,
$\bar{\phi}_I = ( \bar{Q}_i, \bar{B}_\alpha, \bar{C}_{\alpha b} )$ the 
extended action $S_{\rm ext}(A| \phi, \phi_a^*, \bar{\phi})$ is 
required to obey the quantum master equations of the (anti)BRST--symmetry 
together with the type--I Ward identity, both being properly extended to the 
additional antifields.

If this approach is extended to the $osp(1,2)$--symmetric Lagrangian 
quantization \cite{5} the corresponding gauge fixed action 
$S_{m, {\rm eff}}(A| \phi)$ additionally depends on a mass parameter $m$ and 
is required to be $osp(1,2)$-- as well as type--I invariant; the extended 
action $S_{m, {\rm ext}}(A| \phi, \phi_a^*, \bar{\phi}, \eta)$ is required 
to satisfy the quantum master equations of both the (anti)BRST-- 
and $Sp(2)$--symmetry. Thereby, in comparison with the $Sp(2)$--approach, an 
additional source $\eta_I = ( D_i, E_\alpha, F_{\alpha b} )$ has to be 
included in order to fulfil the $osp(1,2)$--superalgebra for $m \neq 0$ 
(for basic definitions see Section 2 below).

Whereas the requirement of gauge covariance with respect to $A^i$, 
i.e.,~type--I symmetry, is completely independent of the master equations,
i.e.,~on the postulated (anti)BRST--symmetries, the type--II
and type--III Ward identities are related to the master equations,
i.e.,~they are {\it consistency conditions} of the theory. Indeed, both 
identities can be derived from the quantum master equations by imposing the 
{\it equations of motion} for the auxiliary fields $B^\alpha$ and the 
(anti)ghosts $C^{\alpha b}$. Even more, if 
$S_{m, {\rm ext}}(A| \phi, \phi_a^*, \bar{\phi}, \eta)$ is 
constructed in such a way that for $A^i \neq 0$ the type--I and for $A^i = 0$ 
the type--II (or type--III) Ward identity is satisfied, then its 
$A^i$--{\it dependence is completely determined by these symmetry 
requirements!} Using this the vertex functional of massive irreducible 
gauge theories with background field $A^i$, 
$\Gamma_m(A| \phi, \phi_a^*, \bar{\phi}, \eta)$, can be related to the vertex 
functional without background field,
\begin{equation*}
\Gamma_m(A| \phi, \phi_a^*, \bar{\phi}, \eta) = 
O_m(A| {\delta}/{\delta \phi}, {\delta}/{\delta \bar{\phi}} ) 
\cdot \Gamma_m(0| \phi, \phi_a^*, \bar{\phi}, \eta),
\end{equation*}
by a well defined operation 
$O_m(A| \delta/\delta \phi, \delta/\delta \bar{\phi})$. In this way the
{\em determination of the background dependence is reduced to quantities 
being completely independent of $A^i$}! An analogous relation 
holds for the generating functional of Green's functions 
$Z_m(A| j, \phi_a^*, \bar{\phi}, \eta)$ where 
$j_I = ( J_i, K_\alpha, L_{\alpha b} )$. 
Within the $osp(1,2)$--symmetric formalism it was also possible to give a 
general solution of a problem which has been posed by Rouet \cite{14} but 
solved by him only for the physical Green's functions: There exists a 
relationship between the Green's functions with a generic background field and 
without them. Furthermore, it has been proven that 
the introduction of a background gauge does not change the ultraviolet 
asymptotics of the theory. This is by no means a trivial result since 
$S_{m, {\rm ext}}(A| \phi, \phi_a^*, \bar{\phi}, \eta)$ in general
depends {\em nonlinear} on the antifields  and, therefore, the antifields 
never more may be interpreted as sources of the (anti)BRST transforms of the 
corresponding fields.

The paper is organized as follows. In Section 2 the $osp(1,2)$--symmetric
Lagrangian quantization is shortly reviewed. In Section 3 it is shown
how, starting from a proper solution of the quantum master equations, by
generalized canonical transformations any admissible solution of the master 
equations may be obtained. Section 4 is devoted to the construction of a 
proper solution of the {\em classical} master equations in the presence of  
generic background fields. This solution 
$\bar{S}^{(0)}_{m, {\rm ext}}(A| \phi, \phi_a^*, \bar{\phi}, \eta)$, which is
obtained 
by assuming that it depends only {\it linearly} on the antifields, after 
choosing a gauge obeys all the symmetry requirements of the classical theory 
$S_{\rm cl}(A + Q)$ we started from. In Section 5, again making use of 
generalized canonical transformations, the general solution of the classical 
master equations, 
$S_{m, {\rm ext}}^{(0)}(A| \phi, \phi_a^*, \bar{\phi}, \eta)$, will be 
obtained which, however, now depends {\it nonlinear} on the antifields. It is 
also shown that this solution is stable against small perturbations and, 
imposing the equations of motion of the auxiliary field $B^\alpha$, that it 
depends on three independent parameters ($z$--factors) only. In Section 6 the 
Ward identities for the generating functional 
$\Gamma_m(A| \phi, \phi_a^*, \bar{\phi}, \eta)$ of the 
one--particle--irreducible (1PI) vertex functions are derived. Besides
the generalizations of the Slavnov--Taylor and the Delduc--Sorella
identities we derive the Kluberg-Stern--Zuber and the Lee identity 
which are related to
type--I and type--II symmetry, respectively. In Section 7,
combining both the last two identities, the $A^i$--dependence of
$\Gamma_m(A| \phi, \phi_a^*, \bar{\phi}, \eta)$  is determined. In addition, 
the Ward identity related to type--III symmetry is derived and 
$\Gamma_m(A| \phi, \phi_a^*, \bar{\phi}, \eta)$ is shown to be type--III
invariant. As an application Rouet's result \cite{14} concerning physical 
Green's functions is proven independently. Finally, in Section 8 the
renormalization group equation for 
$\Gamma_m(A| \phi, \phi_a^*, \bar{\phi}, \eta)$ is derived and a rigorous
proof is given that the $\beta$--function and the anomalous dimensions are
independent of the background  also in this very general setting.


\section{osp(1,2)--covariant quantization}
\renewcommand{\theequation}{\thesection.\arabic{equation}}
\setcounter{equation}{0} 


Now we shortly review the $osp(1,2)$--symmetric quantization procedure 
\cite{5}. Let us start with the widely known $Sp(2)$--covariant formalism 
\cite{3}. It is characterized by introducing to each field
$\phi^I = (Q^i, B^\alpha, C^{\alpha b})$ with Grassmann parity
$\epsilon(\phi^I) \equiv \epsilon_I$, three kinds of antifields,
$\phi_{I a}^* = ( Q_{i a}^*, B_{\alpha a}^*, C_{\alpha ab}^*)$,
$\epsilon(\phi_{I a}^*) = \epsilon_I + 1$ and
$\bar{\phi}_I = ( \bar{Q}_i, \bar{B}_\alpha, \bar{C}_{\alpha b})$,
$\epsilon(\bar{\phi}_I) = \epsilon_I$. Here $\phi_{I a}^*$ 
correspond to (the negative of) the sources of the (anti)BRST- 
transforms, $\mathbf{s}^a \phi^I$, while $\bar{\phi}_I$ corresponds to the 
(negative of the) source of their combined transforms,
$\epsilon_{ab} \mathbf{s}^b \mathbf{s}^a \phi^I$; the index $a= 1,2$ 
labels the $Sp(2)$--doublets of the (anti)BRST--operators $\mathbf{s}^a$ as 
well as of the (anti)ghosts fields. Raising and lowering of $Sp(2)$--indices 
is obtained by the invariant antisymmetric tensor
\begin{equation*}
\epsilon^{ab} = \begin{pmatrix} 0 & 1 \\ -1 & 0 \end{pmatrix},
\qquad
\epsilon^{ac} \epsilon_{cb} = \delta^a_b.
\end{equation*}
 
In the $Sp(2)$--approach a doublet of odd graded symplectic structures,
the extended antibrackets $(F,G)^a$, is introduced by
\begin{equation}
\label{2.1}
(F,G)^a = \frac{\delta F}{\delta \phi^I} \frac{\delta G}{\delta \phi_{I a}^*}
- (-1)^{(\epsilon(F) + 1)(\epsilon(G) + 1)} (F \leftrightarrow G),
\end{equation}
and a doublet of odd graded, nilpotent generating operators is defined through
\begin{equation}
\label{2.2}
\bar{\Delta}^a = \Delta^a - (i/\hbar) V^a
\quad 
\hbox{with}
\quad
\Delta^a = (-1)^{\epsilon_I} \frac{\delta_L}{\delta \phi^I}
\frac{\delta}{\delta \phi_{I a}^*}
\quad
\hbox{and}
\quad
V^a = \epsilon^{ab} \phi_{I b}^* \frac{\delta}{\delta \bar{\phi}_I}.
\end{equation}
(Here, we remind the reader being familiar with the $Sp(2)$--approach that 
according to our changement in the definition of the antifields a minus sign 
appears in the definition of $\bar{\Delta}^a$ above; the same occurs at 
various other places).
Let us remark that the odd graded antibrackets may be defined alternatively
by the Leibniz rule for $\Delta^a$ according to
\begin{equation*}
\Delta^a (FG) = (\Delta^a F) G + 
(-1)^{\epsilon(F)} \big( (F,G)^a + F (\Delta^a G) \big).
\end{equation*}
The relative nilpotency of the 
(anti)BRST--operators, $\{ \mathbf{s}^a, \mathbf{s}^b \} = 0$, repeats itself 
in the important relations $\{ \bar{\Delta}^a, \bar{\Delta}^b \} = 0$.

The quantum action $S(\phi, \phi_a^*, \bar{\phi})$ is required to satisfy 
the {\it quantum} master equations:
\begin{equation}
\label{2.3}
\hbox{$\frac{1}{2}$} (S,S)^a - V^a S = i \hbar \Delta^a S
\qquad
\Longleftrightarrow
\qquad
\bar{\Delta}^a {\rm exp}\{ (i/\hbar) S \} = 0
\end{equation}
with the {\em boundary condition} $S|_{\phi_a^* = \bar{\phi} = \hbar = 0} =
S_{\rm cl}(Q)$. 
Of course, the solution of the master equations is not unique. Analogous 
to the case of the BV--approach \cite{2} two proper solutions of (\ref{2.3})
are related by a generalized canonical transformation with an appropriate 
generating functional \cite{4}.
Therefore, the degeneracy of the action $S$ may be removed by  choosing a
sufficiently general gauge--fixing bosonic functional $F(\phi)$, thus 
defining the extended action $S_{\rm ext}(\phi, \phi_a^*, \bar{\phi})$ 
which by construction also satisfies Eq.~(\ref{2.3}):
\begin{equation}
\label{2.4}
{\rm exp}\{ (i/\hbar) S_{\rm ext} \} =
{\rm exp}\{ - i \hbar \hat{T}(F) \}\, {\rm exp}\{ (i/\hbar) S \}
\quad
\hbox{with}
\quad
\hat{T}(F) = \hbox{$\frac{1}{2}$} \epsilon_{ab} \{
\bar{\Delta}^b, [ \bar{\Delta}^a, F ] \}.
\end{equation}

This formalism despite appearing manifest $Sp(2)$--symmetric leads to solutions
of the quantum master equations which also may be $Sp(2)$--nonsymmetric.
The reason for this can be traced back to the fact that the general 
transformation properties of the solutions of (\ref{2.3}) do not restrict $F$ 
to be a $Sp(2)$--scalar. Therefore, in order to ensure $Sp(2)$--symmetry the 
extended quantum action has to be subjected to further requirements.
In addition, the formalism may be generalized to contain also {\it massive}
gauge and (anti)ghost fields, which are necessary at least intermediately
in the BPHZL renormalization procedure to avoid unwanted infrared 
singularities. Of course, then the 
(anti)BRST--transformations, and the operators $\bar{\Delta}^a$, must be 
extended to include also mass terms.

Let us now state the essential modifications of the $Sp(2)$--formalism 
leading to the $osp(1,2)$--symmetric quantization of irreducible
gauge theories with massive fields. First, the linear part of the 
operators (\ref{2.2}) has to be extended to observe the $m$--dependence, 
\begin{equation}
\label{2.5}
\bar{\Delta}_m^a = \Delta^a - (i/\hbar) V_m^a
\end{equation}
with
\vspace*{-.5cm}
\begin{align}
\label{2.6}
\Delta^a &= (-1)^{\epsilon_I} \frac{\delta_L}{\delta \phi^I}
\frac{\delta}{\delta \phi_{I a}^*}
\\
\label{2.7}
V_m^a &= \epsilon^{ab} \phi_{I b}^* \frac{\delta}{\delta \bar{\phi}_I} -
\eta_I \frac{\delta}{\delta \phi_{I a}^*} +
m^2 (P_+)_{I b}^{J a} \bar{\phi}_J \frac{\delta}{\delta \phi_{I b}^*} -
m^2 \epsilon^{ab} (P_-)_{I b}^{J c} \phi_{J c}^* 
\frac{\delta}{\delta \eta_I}, 
\end{align}
and, in addition, three even graded, second order differential operators
$\bar{\Delta}_A$, $A = (0,\pm)$, generating the symplectic group, have to be
introduced,
\begin{equation}
\label{2.8}
\bar{\Delta}_A = \Delta_A - (i/\hbar) V_A
\end{equation}
with
\vspace*{-.5cm}
\begin{align}
\label{2.9}
\Delta_A &= (-1)^{\epsilon_I} (\sigma_A)_I^{~~\!J}
\frac{\delta_L}{\delta \phi^J} \frac{\delta}{\delta \eta_I}
\\
\label{2.10}
V_A &= \bar{\phi}_J (\sigma_A)^J_{~I} \frac{\delta}{\delta \bar{\phi}_I} +
\big(
\phi_{I b}^* (\sigma_A)^b_{~a} + \phi_{J a}^* (\sigma_A)^J_{~I} \big)
\frac{\delta}{\delta \phi_{I a}^*} +
\eta_J (\sigma_A)^J_{~I} \frac{\delta}{\delta \eta_I}.
\end{align}
Here, the following abbreviations have been used:
\begin{equation*}
(P_-)_{I b}^{J a} \equiv (P_+)_{I b}^{J a} +
\big( \delta_I^J - (P_+)_{I c}^{J c} \big) \delta_a^b,
\qquad
(\sigma_A)^J_{~I} \equiv (\sigma_A)^b_{~a} (P_+)_{I b}^{J a},
\end{equation*}
where, for irreducible gauge theories, the matrix $(P_+)_{I b}^{J a}$ is
defined by \cite{5}
\begin{equation*}
(P_+)_{I b}^{J a} \equiv \begin{cases} 
\delta^i_j \delta_b^a & \text{for $I = i$, $J = j$,} 
\\
\delta_\alpha^\beta \delta_b^a & \text{for $I = \alpha$, $J = \beta$,}
\\
\delta_\alpha^\beta (\delta_c^d \delta_b^a + \delta_c^a \delta_b^d) & 
\text{for $I = \alpha c$, $J = \beta d$,}
\\
0 & \text{otherwise~.} \end{cases}
\end{equation*}
Obviously, the operator $P_+$ projects -- up to a factor of 2 for the 
(anti)ghost components -- onto the nontrivial representations of the 
symplectic group. Notice that the linear parts $V_m^a$ and $V_A$ depend and 
act only on the antifields $\phi_{I a}^*$, $\bar{\phi}_I$ and $\eta_I$. 

The operators $\bar{\Delta}_m^a$ and $\bar{\Delta}_A$ generate a Lie 
superalgebra isomorphic to $osp(1,2)$ \cite{15}:
\begin{align}  
\label{2.11}
[ \bar{\Delta}_A, \bar{\Delta}_B ] &= - (i/\hbar)
\epsilon_{AB}^{~~~~\!C} \bar{\Delta}_C, 
\\
\label{2.12}
[ \bar{\Delta}_A, \bar{\Delta}_m^a ] &= - (i/\hbar)
\bar{\Delta}_m^b (\sigma_A)_b^{~a},
\\
\label{2.13}
\{ \bar{\Delta}_m^a, \bar{\Delta}_m^b \} &= (i/\hbar) m^2
(\sigma^A)^{ab} \bar{\Delta}_A;
\end{align}
here, the matrices $\sigma_A$ generate the algebra $sl(2, R)$,
\begin{equation*}
(\sigma_A)_a^{~c} (\sigma_B)_c^{~b} = g_{AB} \delta_a^b + 
\hbox{$\frac{1}{2}$} \epsilon_{ABC} (\sigma^C)_a^{~b},
\end{equation*}
with Cartan metric
\vspace*{-.5cm}
\begin{equation*}
g^{AB} = \begin{pmatrix} 1 & 0 & 0 \\  0 & 0 & 2 \\ 0 & 2 & 0 
\end{pmatrix},
\qquad
g^{AC} g_{CB} = \delta^A_B,
\end{equation*}
and $\epsilon_{ABC}$ is the antisymmetric tensor, $\epsilon_{0+-} = 1$.
The spinorial indices $a,b$ of $\sigma_A$ are raised or lowered according to
\begin{equation*}
(\sigma_A)^{ab} = \epsilon^{ac} (\sigma_A)_c^{~b} = 
(\sigma_A)^a_{~c} \epsilon^{cb} = 
\epsilon^{ac} (\sigma_A)_{cd} \epsilon^{db},
\qquad
(\sigma_A)_a^{~b} = - (\sigma_A)^b_{~a}.
\end{equation*}
Notice, that $sl(2,R)$, the even part of $osp(1,2)$, is isomorphic to 
$sp(2,R)$. For $\sigma_A$ we choose the representation 
$(\sigma_0)_a^{~b} = \tau_3$ and 
$(\sigma_\pm)_a^{~b} = - \hbox{$\frac{1}{2}$} (\tau_1 \pm i \tau_2)$, 
where $\tau_A$ ($A = 1,2,3$) are the Pauli matrices.
Obviously, as long as $m \neq 0$ the operators $\bar{\Delta}_m^a$
are neither nilpotent nor do they anticommute among themselves.

In addition to the $m$--{\it extended} quantum master equations 
\begin{equation}
\label{2.14}
\hbox{$\frac{1}{2}$} (S_m,S_m)^a - V_m^a S = i \hbar \Delta^a S_m
\qquad
\Longleftrightarrow
\qquad
\bar{\Delta}_m^a \,{\rm exp}\{ (i/\hbar) S_m \} = 0,
\end{equation}
which ensure (anti)BRST- invariance, the mass dependent quantum action 
$S_m(\phi, \phi_a^*, \bar{\phi}, \eta)$ is required to satisfy the 
{\em symplectic} master equations which ensure $Sp(2)$--invariance:
\begin{equation}
\label{2.15}
\hbox{$\frac{1}{2}$} \{ S_m,S_m \}_A - V_A S_m = i \hbar \Delta_A S_m
\qquad
\Longleftrightarrow
\qquad
\bar{\Delta}_A \,{\rm exp}\{ (i/\hbar) S_m \} = 0,
\end{equation}
where the {\it new, even graded brackets} are defined by
\begin{equation}
\label{2.16}
\{ F,G \}_A = (\sigma_A)_J^{~~\!I} 
\frac{\delta F}{\delta \phi^I} \frac{\delta G}{\delta \eta_J} + 
(-1)^{\epsilon(F) \epsilon(G)} (F \leftrightarrow G);
\end{equation}
alternatively, these brackets again may be defined by the Leibniz rule for
$\Delta_A$ :
\begin{equation*}
\Delta_A (FG) = (\Delta_A F) G + \{ F,G \}_A + F (\Delta_A G).
\end{equation*}
In the limit $\hbar \rightarrow 0$ from Eqs.~(\ref{2.14}) and (\ref{2.15}) the 
{\it classical} master equations are obtained.

From Eqs.~(\ref{2.13}) and (\ref{2.15}) it follows that, if and only if 
the action $S_m$ is $Sp(2)$--invariant, it can be (anti)BRST--invariant as 
well. Moreover, in order to express the algebra (\ref{2.11}) -- (\ref{2.13}) 
by {\it operator identities}, and to decompose the set of antifields into
linear spaces being {\it irreducible} under the $osp(1,2)$--superalgebra,
one is forced to enlarge the set of  antifields by additional sources
$\eta_I = (D_i, E_\alpha, F_{\alpha b})$ with $\epsilon(\eta_I) = \epsilon_I$.

In order remove the gauge arbitrariness, analogous to Eq. (\ref{2.4}),
the $m$--extended quantum action
$S_{m, {\rm ext}}(\phi, \phi_a^*, \bar{\phi}, \eta)$ will be introduced
according to 
\begin{gather}
\label{2.17}
{\rm exp}\{ (i/\hbar) S_{m, {\rm ext}} \} =
\hat{U}_m(F)\, {\rm exp}\{ (i/\hbar) S_m \},
\\
\hat{U}_m(F) = {\rm exp}\{ - i \hbar \hat{T}_m(F) \}
\quad
\hbox{with}
\quad
\hat{T}_m(F) = \hbox{$\frac{1}{2}$} \epsilon_{ab} \{
\bar{\Delta}_m^b, [ \bar{\Delta}_m^a, F ] \} + (i/\hbar)^2 m^2 F.
\nonumber
\end{gather}
Written explicitly the operator $\hat{T}_m(F)$ becomes
\begin{equation}
\hspace{-.5cm}
\label{2.18}
\hat{T}_m(F) = - \frac{\delta F}{\delta \phi^I} \Big(
\frac{\delta}{\delta \bar{\phi}_I} - 
\hbox{$\frac{1}{2}$} m^2 (P_-)_{Jc}^{Ic} \frac{\delta}{\delta \eta_J} \Big) -
\hbox{$\frac{1}{2}$}(\hbar/i) \epsilon_{ab}
\frac{\delta}{\delta \phi_{I a}^*}
\frac{\delta^2 F}{\delta \phi^I \delta \phi^J} 
\frac{\delta}{\delta \phi_{J b}^*} + (i/\hbar) m^2 F.
\end{equation}
If the gauge fixing functional $F(\phi)$ is chosen as a $Sp(2)$--scalar, and 
if $S_m$ is restricted to depend on $\eta_I$ only {\it linearly}, namely
$\delta S_m/\delta \eta_I = - \phi^I$, then the following relations may 
be shown to hold:
$[ \bar{\Delta}_m^a, \hat{U}_m(F) ]\, {\rm exp}\{ (i/\hbar) S_m \} = 0$ and
$[ \bar{\Delta}_A, \hat{U}_m(F) ]\, {\rm exp}\{ (i/\hbar) S_m \} = 0$. Hence, 
because $S_m$ being a proper solution of the master equations, the extended 
action $S_{m, {\rm ext}}$ satisfies the quantum master equations 
(\ref{2.14}) and (\ref{2.15}) as well. A more detailed discussion 
of these aspects is given in the next Section. 

Before going on we present the explicit expressions for the operators
$\Delta^a$, $V_m^a$ and $\Delta_A$, $V_A$ as well as for the odd and even
bracket structures $(F,G)^a$ and $\{ F,G \}_A$. Thereby we restrict ourselves
to the case of irreducible gauge theories of first rank with semi--simple 
gauge group and bosonic gauge fields, 
$\epsilon_I = (\epsilon_i, \epsilon_\alpha, \epsilon_\alpha + 1)$ with
$\epsilon_i = \epsilon_\alpha = 0$; in that case the components 
$D_i$ and $E_\alpha$ of $\eta_I$ may be choosen equal to zero:
\begin{align}
\label{2.19}
\Delta^a &= \frac{\delta_L}{\delta Q^i} \frac{\delta}{\delta Q_{i a}^*} +
\frac{\delta_L}{\delta B^\alpha} \frac{\delta}{\delta B_{\alpha a}^*} -
\frac{\delta_L}{\delta C^{\alpha b}} \frac{\delta}{\delta C_{\alpha ab}^*},
\\
V_m^a &= \epsilon^{ab} Q_{i b}^* \frac{\delta}{\delta \bar{Q}_i} +
m^2 \bar{Q}_i \frac{\delta}{\delta Q_{i a}^*} +
\epsilon^{ab} B_{\alpha b}^* \frac{\delta}{\delta \bar{B}_\alpha} +
m^2 \bar{B}_\alpha \frac{\delta}{\delta B_{\alpha a}^*} +
\epsilon^{ab} C_{\alpha bc}^* \frac{\delta}{\delta \bar{C}_{\alpha c}} 
\nonumber
\\
\label{2.20}
&\quad~ + m^2 \bar{C}_{\alpha c} ( 
\frac{\delta}{\delta C_{\alpha ac}^*} + 
\frac{\delta}{\delta C_{\alpha ca}^*} ) -
F_{\alpha c} \frac{\delta}{\delta C_{\alpha ac}^*} +
m^2 \epsilon^{ab} ( C_{\alpha bc}^* - C_{\alpha cb}^* )
\frac{\delta}{\delta F_{\alpha c}},
\\
\label{2.21}
\Delta_A &= - (\sigma_A)_a^{~b} 
\frac{\delta_L}{\delta C^{\alpha b}} \frac{\delta}{\delta F_{\alpha a}},
\\
V_A &= Q_{i a}^* (\sigma_A)^a_{~b} \frac{\delta}{\delta \bar{Q}_{i b}^*} +
B_{\alpha a}^* (\sigma_A)^a_{~b} \frac{\delta}{\delta B_{\alpha b}^*} +
\bar{C}_{\alpha a} (\sigma_A)^a_{~b} \frac{\delta}{\delta \bar{C}_{\alpha b}}
\nonumber
\\
\label{2.22}
&\quad~ + \bigr( 
C_{\alpha ac}^* (\sigma_A)^a_{~b} + C_{\alpha ba}^* (\sigma_A)^a_{~c} \bigr)
\frac{\delta}{\delta C_{\alpha bc}^*} +
F_{\alpha a} (\sigma_A)^a_{~b} \frac{\delta}{\delta F_{\alpha b}}
\end{align}
and
\begin{align}
\label{2.23}
(F,G)^a &= \frac{\delta F}{\delta Q^i} \frac{\delta G}{\delta Q_{i a}^*} +
\frac{\delta F}{\delta B^\alpha} \frac{\delta G}{\delta B_{\alpha a}^*} +
\frac{\delta F}{\delta C^{\alpha b}} \frac{\delta G}{\delta C_{\alpha ab}^*} -
(-1)^{(\epsilon(F) + 1) (\epsilon(G) + 1)} (F \leftrightarrow G), 
\\
\label{2.24}
\{ F,G \}_A &= (\sigma_A)_a^{~b} 
\frac{\delta F}{\delta C^{\alpha b}} \frac{\delta G}{\delta F_{\alpha a}} +
(-1)^{\epsilon(F) \epsilon(G)} (F \leftrightarrow G). 
\end{align}


\section{Generalized canonical transformations}
\renewcommand{\theequation}{\thesection.\arabic{equation}}
\setcounter{equation}{0} 


As became obvious from the considerations of the last Section there exists
a whole set of proper solutions of the master equations. In the 
BV--approach a solution $S$ of the master equation is called proper  if
(a) it has a stationary point where the variations with respect to the fields
and the antifields vanish and (b) its Hessian with respect to the fields
and antifields at that stationary point is nonsingular. It has been shown 
that two proper solutions of the master equations are connected by a 
canonical transformation \cite{2}. These notions have a natural
generalization to the $Sp(2)$--covariant approach --  where a corresponding
connection between two proper solutions already has been proven  
\cite{4} -- as well as to the $osp(1,2)$--symmetric quantization procedure.

In this Section we show how, starting from a proper solution of the quantum 
master equations (\ref{2.14}) and (\ref{2.15}), by generalized canonical 
transformations (see Eq.~(\ref{3.13}) below) any admissible solution of 
the same equations may be obtained. The organization of the proof 
-- as well as the resulting solution -- is quite similar to that presented 
in Ref.~\cite{4}. Thereby, it is very remarkable that there exists a formal 
extension of these results (up to mass terms) to the case of 
$osp(1,2)$--symmetry. Actually, we are only interested in generalized
canonical transformation of the classical master equations. However, in order 
to get such a transformation we solve this problem quite generally, 
i.e.,~first we study transformations that allow one to consider the 
characteristic arbitrariness of a solution of the {\it quantum} master 
equations. Then, by taking the limit $\hbar \rightarrow 0$ we recover the 
transformation we are looking for. Finally, we discuss how in this way the 
general solution of the classical master equations may be obtained 
(the explicit construction of this solution will be given in 
Sections 4 and 5).  

To begin with let us assume (a) that two {\it arbitrary} proper solutions, 
$S_m(0)$ and $S_m(1)$, of the quantum master equations are given and (b) that 
a continuous path in the manifold of solutions $S_m(\zeta), 
0 \leq \zeta \leq 1$, i.e., an interpolating functional, exists which 
connects them. Such a functional, for every value of $\zeta$, 
is obliged to satisfy the generating equations
\begin{equation}
\label{3.2}
\hbox{\large $\frac{1}{2}$} ( S_m(\zeta), S_m(\zeta) )^a -
i \hbar \bar{\Delta}_m^a S_m(\zeta) = 0,
\qquad
\hbox{\large $\frac{1}{2}$} \{ S_m(\zeta), S_m(\zeta) \}_A -
i \hbar \bar{\Delta}_A S_m(\zeta) = 0.
\end{equation}
Differentiating these equations with respect to $\zeta$, then for the derivation 
$\partial S_m(\zeta)/\partial \zeta $ one gets the following consistency 
conditions:
\begin{equation}
\label{3.3}
\mathbf{Q}_m^a(\zeta) 
\frac{\partial S_m(\zeta)}{\partial \zeta} = 0,
\qquad
\mathbf{Q}_A(\zeta) 
\frac{\partial S_m(\zeta)}{\partial \zeta} = 0;
\end{equation}
here the operators $\mathbf{Q}_m^a(\zeta)$ and $\mathbf{Q}_A(\zeta)$ 
depend explicitly  on $S_m(\zeta)$ and their action on an arbitrary 
functional $X$ is defined according to
\begin{equation}
\label{3.4}
\mathbf{Q}_m^a(\zeta) X \equiv ( S_m(\zeta), X )^a - 
i \hbar \bar{\Delta}^a_m X,
\qquad
\mathbf{Q}_A(\zeta) X \equiv \{ S_m(\zeta), X \}_A -
i \hbar \bar{\Delta}_A X.
\end{equation}
These operators, like $\bar{\Delta}_m^a$ and $\bar{\Delta}_A $, are a 
realization of the $osp(1,2)$--superalgebra: 
\begin{align}
\label{3.5}
[ \mathbf{Q}_A(\zeta), \mathbf{Q}_B(\zeta) ] &= 
- \epsilon_{AB}^{~~~~\!C} \mathbf{Q}_C(\zeta),
\nonumber
\\
[ \mathbf{Q}_A(\zeta), \mathbf{Q}_m^a(\zeta) ] &= 
- \mathbf{Q}_m^b(\zeta) (\sigma_A)_b^{~a},
\\
\{ \mathbf{Q}_m^a(\zeta), \mathbf{Q}_m^b(\zeta) \} &= 
m^2 (\sigma^A)^{ab} \mathbf{Q}_A(\zeta).
\nonumber
\end{align}

In order to find an explicit expression for the interpolating functional 
$S_m(\zeta)$ subjected to the consistency conditions (\ref{3.3})
let us make for $\partial S_m(\zeta)/\partial \zeta$ the following ansatz: 
\begin{equation}
\label{3.6}
\frac{\partial S_m(\zeta)}{\partial \zeta} = \hat{W}_m(\zeta) Y
\quad
{\rm with}
\quad
\hat{W}_m(\zeta) \equiv \hbox{$\frac{1}{2}$} \epsilon_{ab} 
\mathbf{Q}_m^b(\zeta) \mathbf{Q}_m^a(\zeta) + m^2 ,
\end{equation}
$Y = Y(\phi, \phi_a^*, \bar{\phi}, \eta)$ being an arbitrary {\it local} 
$Sp(2)$--symmetric functional, i.e.,
\begin{equation}
\label{3.7}
(\sigma_A)_J^{~~\!I} \frac{\delta Y}{\delta \phi^I} \phi^J + V_A Y = 0.
\end{equation}
The justification of that ansatz will be given in Appendix A.

Now, taking into account the $osp(1,2)$--symmetry, Eq.~(\ref{3.5}), by a 
straightforward, but tedious direct computation one obtains  
\begin{equation}
\label{3.8}
\mathbf{Q}_m^a(\zeta) \hat{W}_m(\zeta) = - \hbox{$\frac{1}{2}$} m^2
(\sigma^A)^a_{~b} \mathbf{Q}_m^b(\zeta) \mathbf{Q}_A(\zeta),
\qquad
\mathbf{Q}_A(\zeta) \hat{W}_m(\zeta) = 
\hat{W}_m(\zeta) \mathbf{Q}_A(\zeta),
\end{equation}
and the consistency conditions (\ref{3.3}) are fulfilled provided it holds
\begin{align}
\label{3.9}
\mathbf{Q}_A(\zeta) Y \equiv (\sigma_A)_J^{~~\!I} \bigg\{
\frac{\delta S_m(\zeta)}{\delta \phi^I} \frac{\delta Y}{\delta \eta_J} +
\frac{\delta Y}{\delta \phi^I} \Big(
\frac{\delta S_m(\zeta)}{\delta \eta_J} + \phi^J \Big) -  
i \hbar (-1)^{\epsilon_I} 
\frac{\delta_L}{\delta \phi^I} \frac{\delta Y}{\delta \eta_J} \bigg\} = 0.
\end{align}
Obviously, this equation has a solution which is given by
\begin{equation}
\label{3.10}
\frac{\delta S_m(\zeta)}{\delta \eta_I} + \phi^I = 0,
\qquad
\frac{\delta Y}{\delta \eta_I} = 0, 
\end{equation}
where the first of these equations is a nontrivial condition concerning the 
dependence of $S_m(\zeta)$ on $\eta_I$. However, it is a natural condition. 
This may be seen if Eqs.~(\ref{3.2}) are differentiated with respect to 
$\eta_I$ leading to the following requirements
\begin{equation}
\label{3.11}
\mathbf{Q}_m^a(\zeta) \Big( 
\frac{\delta S_m(\zeta)}{\delta \eta_I} + \phi^I \Big) = 0,
\qquad
\mathbf{Q}_A (\zeta) \Big( 
\frac{\delta S_m(\zeta)}{\delta \eta_I} + \phi^I \Big) = 0.
\end{equation}
Therefore, we are forced to require that $S_m(\zeta)$ is {\it linear} in 
$\eta_I$ and that $Y = Y(\phi, \phi_a^*, \bar{\phi})$ is {\it independent} of 
$\eta_I$. Then, as a consequence of the restrictions (\ref{3.10}), the second 
of the equations (\ref{3.2}) simplifies essentially, 
\begin{equation}
\label{3.12}
(\sigma_A)_J^{~~\!I} \frac{\delta S_m(\zeta)}{\delta \phi^I} \phi^J +
V_A S_m(\zeta) = 0,
\end{equation}
showing that, analogous to Eq.~(\ref{3.7}), the interpolating functional is 
$Sp(2)$--symmetric.

Now, we are left with the problem to integrate Eq.~(\ref{3.6}). The solution 
of that differential equation is given by
\begin{gather}
\label{3.13}
{\rm exp}\{ (i/\hbar) S_m(\zeta) \} = 
\hat{U}_m(\zeta Y) {\rm exp}\{ (i/\hbar) S_m(0) \}, 
\\
\hat{U}_m(\zeta Y) = {\rm exp}\{ (\hbar/i) \hat{T}_m(\zeta Y) \}
\quad
\hbox{with}
\quad
\hat{T}_m(Y) = \hbox{$\frac{1}{2}$} \epsilon_{ab} 
\{ \bar{\Delta}_m^b, [ \bar{\Delta}_m^a, Y ] \} + 
(i/ \hbar)^2 m^2 Y.
\nonumber
\end{gather}
The proof is as follows. First, we rewrite the action of $\mathbf{Q}^a_m$ on 
an arbitrary functional $X$:
\begin{equation*}
\mathbf{Q}^a_m (\zeta) X = \exp\{ - (i/\hbar)S_m(\zeta) \}
(\hbar/i) [ \bar{\Delta}^a_m, X ] \exp\{ (i/\hbar) S_m(\zeta) \}.
\end{equation*}
With the help of this expression the differential equation (\ref{3.6}) 
obtains the following form:
\begin{eqnarray*}
\frac{\partial S_m(\zeta)}{\partial \zeta} = 
{\rm exp}\{ - (i/ \hbar) S_m(\zeta) \} \Big\{  
\hbox{$\frac{1}{2}$} \epsilon_{ab} (\hbar/i)^2 
\{ \bar{\Delta}_m^b, [ \bar{\Delta}_m^a, Y ] \} + m^2 Y \Big\}
{\rm exp}\{ (i/ \hbar) S_m(\zeta) \}.
\end{eqnarray*}
Using this result, and the definition of $\hat{T}_m(Y)$, we obtain
\begin{eqnarray*}
\frac{\partial\, {\rm exp}\{ (i/\hbar) S_m(\zeta) \} }{\partial \zeta} =
{\rm exp}\{ (i/\hbar) S_m(\zeta) \} 
(i/\hbar) \frac{\partial S_m(\zeta)}{\partial \zeta} =
(\hbar/i) \hat{T}_m(Y) \, {\rm exp}\{ (i/\hbar) S_m(\zeta) \}.
\end{eqnarray*}
Because of $\zeta \hat{T}_m(Y) = \hat{T}_m(\zeta Y)$ this finishes the proof 
of Eq.~(\ref{3.13}).

Now, by virtue of
\begin{equation}
\label{3.14}
[ \bar{\Delta}_m^a, - i \hbar \frac{\delta}{\delta \eta_I} + \phi^I ] = 0,
\qquad
[ \bar{\Delta}_A, - i \hbar \frac{\delta}{\delta \eta_I} + \phi^I ] = 0,
\qquad
[ Y, - i \hbar \frac{\delta}{\delta \eta_I} + \phi^I ] = 0,
\end{equation}
from (\ref{3.13}) it follows that imposing condition (\ref{3.10}) together 
with (\ref{3.7}) for $\zeta = 0$ is sufficient to ensure their validity 
also for the whole range $0 < \zeta \leq 1$. This proves that $S_m(\zeta)$, 
as given by Eq.~(\ref{3.13}), is a solution of the quantum master equations, 
i.e.,~it fulfills the first of the equations (\ref{3.2}) as well as 
(\ref{3.12}) for any $\zeta$.  Obviously, performing a generalized 
canonical transformation (\ref{3.13}) of a proper solution $S_m(0)$ by an 
appropriately chosen functional $Y$ appears as the general 
procedure of introducing a gauge (see Eq.~(\ref{2.17})). According to
this result the assumption (b) above has been justified a posteriori.
Therefore, the manifold of proper solutions is connected, i.e.,
any two arbitrary proper solutions may be related through 
Eq.~(\ref{3.13}) in an analogous manner as in the case of the BV and 
the $Sp(2)$--covariant formalism.

If Eq.~(\ref{3.13}) is solved iteratively one gets a series expansion of 
$S_m(\zeta)$ in powers of $\zeta$:
\begin{align}
\label{3.15}
S_m(\zeta) &= \sum_{n = 0}^\infty \zeta^n \overset{(n)}{S}_{\!\!m},
\qquad
\overset{(0)}{S}_{\!\!m} = S_m(0),
\\
(n + 1) \overset{(n + 1)}{S}_{\!\!\!\!\!\!m} &= 
\hbox{\large $\frac{1}{2}$} \epsilon_{ab} \Big\{
\sum_{k = 0}^n ( \overset{(k)}{S}_{\!\!m}, 
( \overset{(n - k)}{S}_{\!\!\!\!\!\!m}, Y )^a )^b +
(\hbar/ i) ( \overset{(n)}{S}_{\!\!m}, \bar{\Delta}_m^a Y )^b 
\nonumber\\
& \quad~ + (\hbar/ i) \bar{\Delta}_m^b ( \overset{(n)}{S}_{\!\!m}, Y )^a +
(\hbar/ i)^2 \delta_{n, 0} \bar{\Delta}_m^b \bar{\Delta}_m^a Y \Big\} +
\delta_{n, 0} m^2 Y,
\qquad
n \geq 0;
\nonumber
\end{align}
this shows that if both $S_m(0)$ and $Y$ are local then $S_m(\zeta)$
is a {\it local} functional as well.

Next, if we are interested in a general solution of the classical master 
equations, which will be the case in the next two Sections, we only have to 
take the limit $\hbar \rightarrow 0$ in the expansion (\ref{3.15}). As a 
result we get a quite nontrivial transformation  which converts a proper 
solution $S_m(0)$ of the {\it classical} master equations into another 
solution $S_m(1)$ of the same equations (for notational simplicity we do not 
distinguish them by an additional index from solutions of the quantum master 
equations)
\begin{align}
\label{3.16}
S_m(\zeta) &= \sum_{n = 0}^\infty \zeta^n \overset{(n)}{S}_{\!\!m},
\qquad
\overset{(0)}{S}_{\!\!m} \equiv S_m(0),
\\
(n + 1) \overset{(n + 1)}{S}_{\!\!\!\!\!\!m} &= 
\hbox{\large $\frac{1}{2}$} \epsilon_{ab} \Bigr\{
\sum_{k = 0}^n ( \overset{(k)}{S}_{\!\!m}, 
( \overset{(n - k)}{S}_{\!\!\!\!\!\!m}, Y )^a )^b -
( \overset{(n)}{S}_{\!\!m}, V_m^a Y )^b 
\nonumber\\
& \quad~ -  V_m^b ( \overset{(n)}{S}_{\!\!m}, Y )^a +
\delta_{n, 0} V_m^b V_m^a Y \Bigr\} +
\delta_{n, 0} m^2 Y,
\qquad
n \geq 0.
\nonumber
\end{align}

Finally, we outline the steps how through the use of two special 
transformations of this kind the general solution of the classical master
equations may be obtained. Let us assume that the {\it general} proper 
solution $S_m(0) \equiv W^{(0)}_m$, being {\it linear} with respect to the 
antifields, has been constructed. In the general case of arbitrary $L$--stage
reducible theories with closed or open algebra for $S_m(0)$ one has to 
choose the general proper solution of the classical master equations with 
vanishing new ghost number, i.e., ${\rm ngh}(S_m(0)) = 0$ \cite{2}. In the 
present case of irreducible theories of first--rank with closed algebra this 
requirement is equivalent to the determination of the general proper solution 
being {\it linear} with respect to the antifields. Then, a more general 
solution, being {\it nonlinear} with respect to the antifields, and 
denoted by $S_m(1) \equiv S_m^{(0)}$, can be constructed by choosing 
\begin{equation}
\label{3.17}
S_m(0) \equiv W^{(0)}_m ~ (\hbox{linear}),
\quad
Y \equiv G
\qquad
\Longrightarrow
\qquad
S_m(1) = S_m^{(0)} ~ (\hbox{nonlinear}), 
\end{equation}
where the most general ansatz for the functional $G(\bar{\phi})$, which has 
mass dimension two in four space--time dimension, is uniquely given by
$G(\bar{\phi})\sim g^{IJ} \bar{\phi}_I \bar{\phi}_J$. Furthermore, using the
transformation (\ref{3.16}) once more, the general solution, being denoted 
by $S_m(1) = S_{m, {\rm ext}}^{(0)}$, is obtained from the previous solution 
$S_m^{(0)}$ by choosing 
\begin{equation}
\label{3.18}
S_m(0) \equiv S_m^{(0)} ~ (\hbox{nonlinear}),
\quad
Y \equiv F
\qquad
\Longrightarrow
\qquad
S_m(1) = S_{m, {\rm ext}}^{(0)} ~ (\hbox{nonlinear}),
\end{equation}
with a gauge--fixing functional $F(\phi) \sim g_{IJ}\phi^I \phi^J$ being the 
most general ansatz with respect to the fields $\phi^I$.
Of course, the solution $S_{m, {\rm ext}}^{(0)}$ could have been constructed 
also by interchanging both steps above, i.e.,~there holds the following 
commutative diagram:
\begin{equation}
\label{3.19}
\begin{CD}
W^{(0)}_{m} @>{Y = G}>> S_m^{(0)} 
\\
@V{Y = F}VV @VV{Y = F}V
\\
W^{(0)}_{m, {\rm ext}} @>>{Y = G}> S_{m, {\rm ext}}^{(0)} 
\end{CD}
\end{equation}
thereby, the solution $S_m(1) = W_{m, {\rm ext}}^{(0)}$ is constructed from
$S_m(0) = W_m^{(0)}$ by choosing 
\begin{equation}
\label{3.20}
S_m(0) \equiv W_m^{(0)} ~ (\hbox{linear}),
\quad
Y \equiv F
\qquad
\Longrightarrow
\qquad
S_m(1) = W_{m, {\rm ext}}^{(0)} ~ (\hbox{linear}).
\end{equation}


\section{Proper Solution of classical master equations}
\renewcommand{\theequation}{\thesection.\arabic{equation}}
\setcounter{equation}{0} 


The general problem of how to construct a solution of the quantum master
equations in the $osp(1,2)$--approach has been solved in Ref.~\cite{5}.
Here we show how to proceed in the case of a linear quantum--background 
splitting of the gauge field. The consecutive steps are the following:
\begin{enumerate}
\item[1.]
Starting with the classical action $S_{\rm cl}(A + Q)$, we first construct a 
proper solution $W^{(0)}_m(A| \phi, \phi_a^*, \bar{\phi}, \eta)$ 
of the {\it classical} master equations,
\begin{equation}
\label{4.1}
\hbox{\large $\frac{1}{2}$} ( W^{(0)}_m, W^{(0)}_m)^a - 
V_m^a W^{(0)}_m = 0,
\qquad
\hbox{\large $\frac{1}{2}$} \{ W^{(0)}_m, W^{(0)}_m \}_A - 
V_A W^{(0)}_m = 0, 
\end{equation}
being {\it linear} in the {\it antifields}.

\item[2.]
In order to remove the gauge degeneracy of that solution 
$W^{(0)}_m(A| \phi, \phi_a^*, \bar{\phi}, \eta)$ a gauge is introduced by an
appropriate generating functional $F(\phi)$ according to Eq.~(\ref{3.13})
in the tree approximation. This defines  the extended action 
$W^{(0)}_{m, {\rm ext}}(A| \phi, \phi_a^*, \bar{\phi}, \eta)$ being a solution 
of the classical  master equations  which, in addition, is invariant under 
type--I as well as (up to mass terms) under type--II and type--III 
transformations.
\end{enumerate}
Despite these nice properties, that solution can not be used
for the construction of the generating functional of vertex functions,
$\Gamma_m(A| \phi, \phi_a^*, \bar{\phi}, \eta)$, since it is not stable
under small perturbations, i.e.,~not all of the counter terms which may 
occure in the process of renormalization could be absorbed by redefining
the independent parameters of the theory. This circumstance is due the fact 
that $\phi^I$ and $\bar{\phi}_I$ mix under renormalization. Therefore, 
in the next Section, we continue our programme by the following steps.
\begin{enumerate}
\item[3.]
The proper solution $W^{(0)}_{m}(A| \phi, \phi_a^*, \bar{\phi}, \eta)$ by 
the help of the extension (\ref{3.17}) will be generalized to a solution 
of the classical master equations, 
$S^{(0)}_m(A| \phi, \phi_a^*, \bar{\phi}, \eta)$, which depends 
{\it nonlinear} on the antifields $\phi_{I a}^*$ and $\bar{\phi}_I$ 
(however, because of the requirement (\ref{3.10}), it must be linear in 
$\eta_I$). 
\item[4.]
Now, fixing the gauge as before by the generating functional $F(\phi)$ we 
arrive at the most general solution of the classical master equations,
$S^{(0)}_{m, {\rm ext}}(A| \phi, \phi_a^*, \bar{\phi}, \eta)$, which
will be the starting point for the solution of the quantum master equations, 
i.e.,~the generalization to any order of perturbation theory. It may be shown 
that $S^{(0)}_{m, {\rm ext}}(A| \phi, \phi_a^*, \bar{\phi}, \eta)$
depends on seven independent parameters and that it is invariant 
(up to mass terms) under type--I, type--II and type--III transformations. 
\end{enumerate}

\smallskip
\noindent (A) {\it Construction of a proper solution}
\\
The symmetry operators of the action $W^{(0)}_m$, the $m$--extended (anti)BRST 
operator and the generators of symplectic transformations, will be 
denoted by $\mathbf{s}_m^a$ ($a = 1,2$) and $\mathbf{d}_A$ ($A = 0,+,-$), 
respectively. They fulfil the $osp(1,2)$--superalgebra in the following form:
\begin{equation}
\label{4.2}
[ \mathbf{d}_A, \mathbf{d}_B ] = 
\epsilon_{ABC} \,\mathbf{d}^C,
\qquad
[ \mathbf{d}_A, \mathbf{s}_m^a ] = 
\mathbf{s}_m^b (\sigma_A)_b^{~a},
\qquad
\{ \mathbf{s}_m^a, \mathbf{s}_m^b \} =
 - m^2 (\sigma^A)^{ab}\, \mathbf{d}_A.
\end{equation}
In order to ensure the $osp(1,2)$--invariance of $W^{(0)}_m$ we make the 
following ansatz:
\begin{equation}
\label{4.3}
W^{(0)}_m = S_{\rm cl}(A + Q) - ( \hbox{$\frac{1}{2}$} \epsilon_{ab} 
\mathbf{s}_m^b \mathbf{s}_m^a + m^2) X 
\quad
\hbox{with}
\quad
X = \bar{Q}_i Q^i + \bar{B}_\alpha B^\alpha + \bar{C}_{\alpha a} C^{\alpha a}.
\end{equation}
Obviously, the term to be added to $S_{\rm cl}(A + Q)$ has a structure which
is analogous to the right hand side of Eq.~(\ref{3.6}). Furthermore, $X$ 
is chosen to be a $Sp(2)$--scalar, $\mathbf{d}_A X = 0$; in fact, it is the 
only one we are able to build up {\it linear} in the antifields. Now, because 
$S_{\rm cl}(A + Q)$ is gauge invariant and since, by virtue of (\ref{4.2}), 
one verifies the equalities 
$\mathbf{s}_m^c ( \hbox{$\frac{1}{2}$} \epsilon_{ab} 
\mathbf{s}_m^b \mathbf{s}_m^a + m^2) X = \hbox{$\frac{1}{2}$} m^2
(\sigma^A)^c_{~d} \mathbf{s}_m^d \mathbf{d}_A X = 0$ and 
$[ \mathbf{d}_A, \hbox{$\frac{1}{2}$} \epsilon_{ab} 
\mathbf{s}_m^b \mathbf{s}_m^a + m^2 ] X = 0$, it easily follows that 
$W_m^{(0)}$ is both (anti)BRST-- and $Sp(2)$--invariant,
\begin{equation*}
\mathbf{s}_m^a W^{(0)}_m = 0,
\qquad
\mathbf{d}_A W^{(0)}_m = 0.
\end{equation*}
For fixed $A^i$ and for the case of irreducible closed gauge algebra, 
the explicit expressions of  the (anti)BRST-- and $Sp(2)$--transformations 
of the fields are uniquely defined by 
\begin{alignat}{4}
\mathbf{s}_m^a A^i~ &= 0,
&\qquad
\mathbf{d}_A A^i~ &= 0,
\nonumber\\
\label{4.4}
\mathbf{s}_m^a Q^i~ &= R^i_\alpha(A + Q) C^{\alpha a},
&\qquad
\mathbf{d}_A Q^i~ &= 0,
\\
\mathbf{s}_m^a C^{\alpha b} &= \epsilon^{ab} B^\alpha -
\hbox{$\frac{1}{2}$} f^\alpha_{\beta\gamma} C^{\beta a} C^{\gamma b},
&\qquad
\mathbf{d}_A C^{\alpha b} &= C^{\alpha d} (\sigma_A)_d^{~b},
\nonumber\\
\mathbf{s}_m^a B^\alpha\, &= 
\hbox{$\frac{1}{2}$} f^\alpha_{\beta\gamma} B^\beta C^{\gamma a} +
\hbox{$\frac{1}{12}$} \epsilon_{cd} f^\alpha_{\eta\beta} f^\eta_{\gamma\delta} 
C^{\gamma a} C^{\delta c} C^{\beta d} - m^2 C^{\alpha a},
&\qquad
\mathbf{d}_A B^\alpha\, &= 0.
\nonumber
\end{alignat}
Here, the (anti)BRST- transformations of $A^i$ and $Q^i$ are obtained from the
type--II transformations (\ref{1.5}) with $\xi^\alpha$ replaced by the 
(anti)ghosts $C^{\alpha a}$. (Let us point to the fact that for irreducible
gauge theories the auxiliary field $B^\alpha$ and the (anti)ghosts 
$C^{\alpha b}$ are related to each other according to 
$B^\alpha = \frac{1}{2} \epsilon_{ab}\mathbf{s}_m^a C^{\alpha b}$.) 
In principle, one could replace also the type--III transformations (\ref{1.6}) 
by defining $\mathbf{s}_m^a A^i = R^i_{\alpha, j} A^j C^{\alpha a}$ and 
$\mathbf{s}_m^a Q^i = R^i_\alpha(Q) C^{\alpha a}$; but, such a definition 
comes into conflict with the background gauge covariance of the theory. 

The corresponding transformations of the antifields, taking into account 
their Grassmann parity and dimension as well as their transformation 
properties under $Sp(2)$, are uniquely given as
\begin{alignat}{4}
\mathbf{s}_m^a \bar{Q}_i~~ &= \epsilon^{ab} Q_{i b}^*,
&\qquad
\mathbf{d}_A \bar{Q}_i~~ &= 0,
\nonumber\\
\mathbf{s}_m^a Q_{i b}^*~\, &= m^2 \delta^a_b \bar{Q}_i,
&\qquad
\mathbf{d}_A Q_{i b}^*~\, &= Q_{i d}^* (\sigma_A)^d_{~b},
\nonumber\\
\mathbf{s}_m^a \bar{B}_\alpha~~ &= \epsilon^{ab} B_{\alpha b}^*,
&\qquad
\mathbf{d}_A \bar{B}_\alpha~~ &= 0,
\nonumber\\
\label{4.5}
\mathbf{s}_m^a B_{\alpha b}^*~ &= m^2 \delta^a_b \bar{B}_\alpha,
&\qquad
\mathbf{d}_A B_{\alpha b}^*~ &= B_{\alpha d}^* (\sigma_A)^d_{~b},
\\
\mathbf{s}_m^a \bar{C}_{\alpha c}~ &= \epsilon^{ab} C_{\alpha bc}^*,
&\qquad
\mathbf{d}_A \bar{C}_{\alpha c}~ &= \bar{C}_{\alpha d} (\sigma_A)^d_{~c},
\nonumber\\
\mathbf{s}_m^a F_{\alpha c}~\, &= m^2 \epsilon^{ab} ( 
C_{\alpha bc}^* - C_{\alpha cb}^* ),
&\qquad
\mathbf{d}_A F_{\alpha c}~\, &= F_{\alpha d} (\sigma_A)^d_{~c},
\nonumber\\
\mathbf{s}_m^a C_{\alpha bc}^* &= m^2 ( 
\delta^a_b \bar{C}_{\alpha c} + \delta^a_c \bar{C}_{\alpha b} ) - 
\delta^a_b F_{\alpha  c},
&\qquad
\mathbf{d}_A C_{\alpha bc}^* &= C_{\alpha dc}^* (\sigma_A)^d_{~b} + 
C_{\alpha bd}^* (\sigma_A)^d_{~c},
\nonumber
\end{alignat}
Using the explicit expressions (\ref{4.4}) and (\ref{4.5}) for the proper 
solution (\ref{4.3}) one obtains
\begin{align}
\label{4.6}
W^{(0)}_m &= S_{\rm cl}(A + Q) - 
( \epsilon^{ab} C^*_{\alpha ba} - m^2 \bar{B}_\alpha ) B^\alpha - 
( F_{\alpha a} - m^2 B^*_{\alpha a} ) C^{\alpha a}
\nonumber\\
&\quad - Q^*_{i a} R^i_\alpha(A + Q) C^{\alpha a} - \bar{Q}_i \bigr(
R^i_\alpha(A + Q) B^\alpha +
\hbox{$\frac{1}{2}$} \epsilon_{ab} R^i_{\alpha, j} 
R^j_\beta(A + Q) C^{\beta b} C^{\alpha a} \bigr)
\\
&\quad + C^*_{\alpha ab} (
- \hbox{$\frac{1}{2}$} f^\alpha_{\beta\gamma} C^{\beta a} C^{\gamma b} ) +
( \bar{C}_{\alpha a} - \hbox{$\frac{1}{2}$} B^*_{\alpha a} ) \bigr(
f^\alpha_{\beta\gamma} B^\beta C^{\gamma a} +
\hbox{$\frac{1}{6}$} \epsilon_{cd} 
f^\alpha_{\eta\beta} f^\eta_{\gamma\delta}
C^{\gamma a} C^{\delta c} C^{\beta d} \bigr).
\nonumber
\end{align}
From this explicit expression for $W^{(0)}_m$ one reads off the relations 
\begin{equation}
\label{4.7}
\hbox{\large $\frac{1}{2}$} \epsilon_{ab}
\frac{\delta}{\delta C^*_{\alpha ba}} W^{(0)}_m = B^\alpha
\quad
\hbox{and}
\quad
\frac{\delta}{\delta F_{\alpha a}} W^{(0)}_m = - C^{\alpha a},
\end{equation}
which imply, by using the master equations (\ref{4.1}), two additional 
conditions for $W^{(0)}_m$,
\begin{equation}
\label{4.8}
\Big( \frac{\delta}{\delta B^*_{\alpha a}} +
\hbox{\large$\frac{1}{2}$} \frac{\delta}{\delta \bar{C}_{\alpha a}} \Big) 
W^{(0)}_m = - m^2 \frac{\delta}{\delta F_{\alpha a}} W^{(0)}_m
\quad
\hbox{and}
\quad
\frac{\delta}{\delta \bar{B}_\alpha} W^{(0)}_m = 
\hbox{\large$\frac{1}{2}$} m^2 \epsilon_{ab}
\frac{\delta}{\delta C^*_{\alpha ba}} W^{(0)}_m.
\end{equation}
Thus, $W^{(0)}_m$ depends on $B^*_{\alpha a}$ and $\bar{B}_\alpha$
only through the combinations
\begin{equation}
\label{4.9}
\bar{E}_{\alpha a} \equiv 
\bar{C}_{\alpha a} - \hbox{$\frac{1}{2}$} B^*_{\alpha a},
\qquad 
E^*_{\alpha ab} \equiv 
C^*_{\alpha ab} - \hbox{$\frac{1}{2}$} m^2 \epsilon_{ab} \bar{B}_\alpha,
\qquad
H_{\alpha a} \equiv F_{\alpha a} - m^2 B^*_{\alpha a},
\end{equation}
respectively, which may be read off also from Eq.~(\ref{4.6}). Of course, 
these combinations transform under $\mathbf{s}_m^a$ and $\mathbf{d}_A$ 
exactly as their first components $\bar{C}_{\alpha a}$, $C_{\alpha ab}^*$ 
and $F_{\alpha a}$, 
\begin{alignat}{4}
\mathbf{s}_m^a \bar{E}_{\alpha c} &= 
\epsilon^{ab} E_{\alpha bc}^*,
&\qquad
\mathbf{d}_A \bar{E}_{\alpha c} &= \bar{E}_{\alpha d} (\sigma_A)^d_{~c},
\nonumber
\\
\label{4.10}
\mathbf{s}_m^a H_{\alpha c} &= m^2 \epsilon^{ab} ( 
E_{\alpha bc}^* - E_{\alpha cb}^* ),
&\qquad
\mathbf{d}_A H_{\alpha c} &= H_{\alpha d} (\sigma_A)^d_{~c},
\\
\mathbf{s}_m^a E_{\alpha bc}^* &= m^2 ( 
\delta^a_b \bar{E}_{\alpha c} + \delta^a_c \bar{E}_{\alpha b} ) - 
\delta^a_b H_{\alpha c},
&\qquad
\mathbf{d}_A E_{\alpha bc}^* &= E_{\alpha dc}^* (\sigma_A)^d_{~b} + 
E_{\alpha bd}^* (\sigma_A)^d_{~c}.
\nonumber
\end{alignat}
Thus, the dependence of $W^{(0)}_m$ on the antifields $B^*_{\alpha b}$ and 
$\bar{B}_\alpha$ is completely determined by the classical master equations 
(\ref{4.1}) together with the relations (\ref{4.7}) and (\ref{4.8}). 
(The $B^\alpha$--dependence is much more involved; it will be considered
later on.) Of course, Eqs.~(\ref{4.10}) replace the last five sets of 
Eqs.~in (\ref{4.5}). This allows to rewrite the action (\ref{4.3}) as
\begin{equation*}
W^{(0)}_m = S_{\rm cl}(A + Q) - ( \hbox{$\frac{1}{2}$} \epsilon_{ab} 
\mathbf{s}_m^b \mathbf{s}_m^a + m^2 ) Y 
\quad
\hbox{with}
\quad
Y = \bar{Q}_i Q^i + \bar{E}_{\alpha a} C^{\alpha a},
\end{equation*}
where the combination
$\bar{B}_\alpha B^\alpha + \bar{C}_{\alpha b} C^{\alpha b}$ 
in the earlier definition of $X$ are now replaced by 
$\bar{E}_{\alpha b} C^{\alpha b}$. Then, for $W^{(0)}_m$ one obtains 
\begin{align}
\label{4.11}
W^{(0)}_m &= S_{\rm cl}(A + Q) - H_{\alpha a} C^{\alpha a}
\nonumber\\
&\quad - Q^*_{i a} R^i_\alpha(A + Q) C^{\alpha a} - \bar{Q}_i \big(
R^i_\alpha(A + Q) B^\alpha +
\hbox{$\frac{1}{2}$} \epsilon_{ab} R^i_{\alpha, j} 
R^j_\beta(A + Q) C^{\beta b} C^{\alpha a} \big)
\\
&\quad + E^*_{\alpha ab} ( \epsilon^{ab} B^\alpha -
\hbox{$\frac{1}{2}$} f^\alpha_{\beta\gamma} C^{\beta a} C^{\gamma b} ) +
\bar{E}_{\alpha a} \big(
f^\alpha_{\beta\gamma} B^\beta C^{\gamma a} +
\hbox{$\frac{1}{6}$} \epsilon_{cd} f^\alpha_{\eta\beta} f^\eta_{\gamma\delta}
C^{\gamma a} C^{\delta c} C^{\beta d} \big) 
\nonumber
\end{align}
which is a proper solution of the classical master equations being regular 
in the neighborhood of the stationary point. The proof that expression
(\ref{4.11}) is the most general solution of the classical master equations
depending only linearly on the antifields is given in Appendix A.

\smallskip
\noindent (B) {\it Gauging of the proper solution}
\\
In order to lift the degeneracy of $W^{(0)}_m$ we still have to introduce a 
gauge. The corresponding extended action $W_{m, {\rm ext}}^{(0)}$ is obtained
from the canonical transformation Eq.~(\ref{3.16}) according to the choice
Eq.~(\ref{3.20}). In the case of linear dependence on the antifields this 
transformation reduces considerably. Namely, choosing a {\it minimal} gauge, 
i.e.,~a gauge which depends only on the independent dynamical fields $Q^i$ and 
$C^{\alpha a}$, the extended action $W_{m, {\rm ext}}^{(0)}$ is obtained 
through the following construction
\begin{equation}
\label{4.12}
W^{(0)}_{m, {\rm ext}} = W^{(0)}_m + (\hbox{$\frac{1}{2}$}
\epsilon_{ab} 
\mathbf{s}_m^b \mathbf{s}_m^a + m^2) F 
\quad
\hbox{with}
\quad
F = \hbox{$\frac{1}{2}$} (
g_{ij} Q^i Q^j + \xi g_{\alpha\beta} \epsilon_{ab} C^{\alpha a} C^{\beta b} ),
\end{equation}
$\xi$ being an arbitrary gauge parameter. Again, the gauge--fixing functional
$F$ is assumed to be a $Sp(2)$--scalar, $\mathbf{d}_A F = 0$. The gauge 
fixing terms in (\ref{4.12}) extend the action $W^{(0)}_m$ according to
\begin{align}
W^{(0)}_{m, {\rm ext}} &= W^{(0)}_m + g_{ij} \bigr( 
Q^i R^j_\alpha(A) B^\alpha + \hbox{$\frac{1}{2}$} \epsilon_{ab}  
R^i_\alpha(A) C^{\alpha b} R^j_\beta(A + Q) C^{\beta a} + 
\hbox{$\frac{1}{2}$} m^2 Q^i Q^j \bigr)
\nonumber
\\
\label{4.13}
&\quad + \xi g_{\alpha\beta} \bigr(
B^\alpha B^\beta -
\hbox{$\frac{1}{24}$} \epsilon_{ab} \epsilon_{cd} 
( f^\alpha_{\gamma\rho} C^{\gamma a} C^{\rho c} )  
( f^\beta_{\delta\sigma} C^{\delta b} C^{\sigma d} ) +
m^2 \epsilon_{ab} C^{\alpha a} C^{\beta b} \bigr), 
\end{align}
where use has been made of the symmetry properties of the structure constants 
and the relations $g_{ki} R^i_{\alpha, j} = - g_{ji} R^i_{\alpha, k}$ and
$g^{ik} R^j_{\alpha, i} = - g^{ij} R^k_{\alpha, i}$. Note, that this 
background gauge is {\it nonlinear} due the occurence of quartic (anti)ghosts 
terms. By construction, $W^{(0)}_{m, {\rm ext}}$ is both (anti)BRST-- and 
$Sp(2)$--invariant,
\begin{equation*}
\mathbf{s}_m^a W^{(0)}_{m, {\rm ext}} = 0,
\qquad 
\mathbf{d}_A W^{(0)}_{m, {\rm ext}} = 0.
\end{equation*}
Hence, it provides also a solution of the Eqs.~(\ref{4.1}) and, in addition, 
it satisfies the constraints implied by the Eqs.~(\ref{4.7}) and (\ref{4.8}).

Furthermore, it can be verified that $W^{(0)}_{m, {\rm ext}}$, in accordance 
with (\ref{1.4}) -- (\ref{1.6}), is invariant under both background gauge 
(type--I) as well as quantum gauge (type--II and type--III) transformations 
(except for the mass term of $Q^i$):
\begin{align}
\label{4.14}
\hbox{type--I}:
\qquad
\delta W^{(0)}_{m, {\rm ext}} &= 0,
\\
\delta Q^i &= R^i_{\alpha, j} Q^j \xi^\alpha,
\qquad
\delta A^i = R^i_\alpha(A) \xi^\alpha,
\qquad
\delta \bar{Q}_i = - R^j_{\alpha, i} \bar{Q}_j \xi^\alpha, ~~ \ldots,
\nonumber\\
\label{4.15}
\hbox{type--II}:
\qquad
\delta W^{(0)}_{m, {\rm ext}} &= m^2 g_{ij} R^i_\alpha(A) Q^j \xi^\alpha,
\\
\delta Q^i &= R^i_\alpha(A + Q) \xi^\alpha,
\qquad
\delta A^i = 0,
\qquad
\delta \bar{Q}_i = g_{ij} R^j_\alpha(A + \bar{Q}) \xi^\alpha, ~~ \ldots,
\nonumber\\
\label{4.16}
\hbox{type--III}:
\qquad
\delta W^{(0)}_{m, {\rm ext}} &= m^2 g_{ij} R^i_\alpha(0) Q^j \xi^\alpha,
\\
\delta Q^i &= R^i_\alpha(Q) \xi^\alpha,
\qquad
\delta A^i = R^i_{\alpha, j} A^j \xi^\alpha,
\qquad
\delta \bar{Q}_i = g_{ij} R^j_\alpha(\bar{Q}) \xi^\alpha, ~~ \ldots;
\nonumber
\end{align}
here the ellipses $\ldots$ indicate the transformations of all the other 
(anti)fields which transform according to the adjoint representation:
\begin{alignat}{3}
&&\qquad
\delta B^\alpha 
&= {\phantom{-}}f^\alpha_{\beta\gamma} B^\beta \xi^\gamma,
&\qquad
\delta C^{\alpha b} 
&= {\phantom{-}}f^\alpha_{\beta\gamma} C^{\beta b} \xi^\gamma,
\nonumber\\
\label{4.17}
\delta F_{\alpha b} &= - f^\beta_{\alpha\gamma} F_{\beta b} \xi^\gamma.
&\qquad
\delta \bar{B}_\alpha &= - f^\beta_{\alpha\gamma} \bar{B}_\beta \xi^\gamma,
&\qquad
\delta \bar{C}_{\alpha b} &= - f^\beta_{\alpha\gamma} \bar{C}_{\beta b}
\xi^\gamma,
\\
\delta Q^*_{i a} &= - R^j_{\alpha, i} Q^*_{j a} \xi^\alpha,
&\qquad
\delta B^*_{\alpha a} &= - f^\beta_{\alpha\gamma} B^*_{\beta a} \xi^\gamma,
&\qquad
\delta C^*_{\alpha ab} &= - f^\beta_{\alpha\gamma} C^*_{\beta ab} \xi^\gamma.
\nonumber
\end{alignat} 
Let us mention, that the symmetries (\ref{4.15}) and (\ref{4.16}) do not hold 
for the effective action $S_{m, {\rm eff}}^{(0)}$ but only for the extended 
action $W_{m, {\rm ext}}^{(0)}$.


\section{General solution of the master equations}
\renewcommand{\theequation}{\thesection.\arabic{equation}}
\setcounter{equation}{0} 


So far, we have ignored the important question whether the action 
(\ref{4.13}), which was supposed to be {\it linear} with respect to the 
antifields, is the general solution of the classical master equations 
(\ref{4.1}) subjected to the conditions (\ref{4.7}) and (\ref{4.8}).
Unfortunately, this is not the case because  $W^{(0)}_m$ is not stable against 
small perturbations. This may be traced back to the fact that the fields 
$Q^i$, $C^{\alpha a}$ and the related antifields $\bar{Q}_i$, 
$\bar{E}_{\alpha a}$ have the same quantum numbers, respectively, and 
therefore mix under renormalization. More precisely, without violating the 
$osp(1,2)$--superalgebra, the (anti)BRST-- and $Sp(2)$--transformations 
(\ref{4.4}) may be altered with $Q^i$, $C^{\alpha a}$ and 
$B^\alpha = \hbox{$\frac{1}{2}$} \epsilon_{ba} \mathbf{s}_m^a C^{\alpha b}$ 
being replaced by
\begin{gather}
\label{5.1}
\tilde{Q}^i \equiv Q^i + \tilde{\sigma} g^{ij} \bar{Q}_j,
\\
\tilde{C}^{\alpha a} \equiv C^{\alpha a} + 
\tilde{\rho} \xi^{-1} g^{\alpha\beta} \epsilon^{ab} \bar{E}_{\beta b}
\quad
\hbox{and}
\quad
\tilde{B}^\alpha \equiv B^\alpha + \hbox{$\frac{1}{2}$} 
\tilde{\rho} \xi^{-1} g^{\alpha\beta} \epsilon^{ab} E^*_{\beta ab},
\nonumber
\end{gather}
respectively, where the following abbreviations are introduced:
\begin{equation*}
\tilde{\sigma} \equiv 1 - \sigma,
\qquad
\tilde{\rho} \equiv 1 - \rho,
\end{equation*} 
$\sigma$ and $\rho$ being independent dimensionless parameters; $\xi^{-1}$ 
has been introduced for later convenience where it will be identified with 
the (inverse) gauge parameter (before introducing a gauge we could use 
$\xi = 1$). Now, in order to ensure stability of the action $W_m^{(0)}$ we 
introduce the antifields in a {\it nonlinear} way by terms which depend on 
$\sigma$ and $\rho$. Of course, if the action depends nonlinearly on the 
antifields these quantities lose their interpretation as sources of the 
(anti)BRST transforms of the fields.

\smallskip
\noindent (A) {\it General solution of the classical master equations 
prior to gauge fixing}
\\
Let us ignore the procedure of generalized canonical transformations as 
introduced in Section 3 and simply change the solution $W_m^{(0)}$ in an 
analogous way as we have done for fixing the gauge, namely by adding the 
following {\it nonlinear} terms 
\begin{equation*}
X_m = W_m^{(0)} + ( \hbox{$\frac{1}{2}$} \epsilon_{ab} 
\mathbf{s}_m^b \mathbf{s}_m^a + m^2 ) G 
\quad
\hbox{with}
\quad
G = \hbox{$\frac{1}{2}$} (
\tilde{\sigma} g^{ij} \bar{Q}_i \bar{Q}_j + 
\tilde{\rho} \xi^{-1} g^{\alpha\beta} \epsilon^{ab} 
\bar{E}_{\alpha a} \bar{E}_{\beta b} ).
\end{equation*}
Carrying out the replacements (\ref{5.1}) in the proper solution (\ref{4.11}) 
and making use of (\ref{4.5}) and (\ref{4.10}) then for the resulting 
functional, $\tilde{X}_m$, one gets
\begin{align*}
\tilde{X}_m &= S_{\rm cl}(A + \tilde{Q}) - H_{\alpha a} ( 
\tilde{C}^{\alpha a} - 
\tilde{\rho} \xi^{-1} g^{\alpha\beta} \epsilon^{ab} \bar{E}_{\beta b} )
\\
&\quad + \hbox{$\frac{1}{2}$} \tilde{\sigma} g^{ij} (
\epsilon^{ab} Q^*_{i a} Q^*_{j b} - m^2 \bar{Q}_i \bar{Q}_j ) -
\tilde{\rho} \xi^{-1} g^{\alpha\beta} \epsilon^{ab} ( 
\hbox{$\frac{1}{2}$} \epsilon^{cd} E^*_{\alpha ac} E^*_{\beta bd} + 
m^2 \bar{E}_{\alpha a} \bar{E}_{\beta b} ) 
\\
&\quad - Q^*_{i a} R^i_\alpha(A + \tilde{Q}) \tilde{C}^{\alpha a} - 
\bar{Q}_i \bigr(
R^i_\alpha(A + \tilde{Q}) \tilde{B}^\alpha +
\hbox{$\frac{1}{2}$} \epsilon_{ab} R^i_{\alpha, j} 
R^j_\beta(A + \tilde{Q}) \tilde{C}^{\beta b} \tilde{C}^{\alpha a} \bigr)
\\
&\quad + E^*_{\alpha ab} (
\epsilon^{ab} \tilde{B}^\alpha - \hbox{$\frac{1}{2}$}  
f^\alpha_{\beta\gamma} \tilde{C}^{\beta a} \tilde{C}^{\gamma b} ) +
\bar{E}_{\alpha a} \bigr(
f^\alpha_{\beta\gamma} \tilde{B}^\beta \tilde{C}^{\gamma a} +
\hbox{$\frac{1}{6}$} \epsilon_{cd} 
f^\alpha_{\eta\beta} f^\eta_{\gamma\delta}
\tilde{C}^{\gamma a} \tilde{C}^{\delta c} \tilde{C}^{\beta d} \bigr).
\end{align*}
By construction $\tilde{X}_m$ is both (anti)BRST-- and $Sp(2)$--invariant,
\begin{equation*}
\mathbf{\tilde{s}}_m^a \tilde{X}_m = 0,
\qquad 
\mathbf{d}_A \tilde{X}_m = 0,
\end{equation*} 
where the action of $\mathbf{\tilde{s}}_m^a$ on the fields is defined by
carring out in (\ref{4.4}) the replacements (\ref{5.1}).  
But, unfortunately, owing to its nonlinear dependence on the antifields it 
does not solve the master equations (\ref{4.1}); instead it holds
\begin{equation*}
\hbox{\large $\frac{1}{2}$} ( \tilde{X}_m, \tilde{X}_m )^a 
- V_m^a \tilde{X}_m = 
\tilde{\rho} \xi^{-1} g^{\alpha\beta} \epsilon^{ab}
\frac{\delta \tilde{X}_m}{\delta B^\alpha}
\frac{\delta \tilde{X}_m}{\delta C^{\beta b}},
\qquad
\hbox{\large $\frac{1}{2}$} \{ \tilde{X}_m, \tilde{X}_m \}_A 
- V_A \tilde{X}_m = 0.
\end{equation*}
The reason for this failure is that we did not really apply the correct 
transformation law, Eq.~(\ref{3.16}). However, the symmetry breaking term 
on the right hand side may be compensated by the following construction
\begin{equation}
\label{5.2}
S_m^{(0)} = \tilde{X}_m +
\hbox{\large $\frac{1}{4}$} \tilde{\rho} \xi^{-1} g^{\alpha\beta} 
\frac{\delta \tilde{X}_m}{\delta B^\alpha}
\frac{\delta \tilde{X}_m}{\delta B^\beta}.
\end{equation}
Of course, by using the more sophisticated transformation law,
Eq.~(\ref{3.16}), with the choice Eq.~(\ref{3.17}) for the generating 
functional $G$, we would be led directly to expression (\ref{5.2}) 
(for a detailed proof see Appendix B).

\smallskip
\noindent (B) {\it Determination of the extended action 
$S_{m, {\rm ext}}^{(0)}$}
\\
Now, we are left with the problem to introduce a gauge. Again, we  choose
the {\it minimal} gauge by the functional $F(\phi)$ since any other term, 
which could be considered as part of a {\it nonminimal} gauge, already have 
been taken into account by the above construction. Analogous to 
Eq.~(\ref{5.2}) it is given by (see Appendix B)
\begin{equation}
\label{5.3}
S_{m, {\rm ext}}^{(0)} = \tilde{Y}_m +
\hbox{\large$\frac{1}{4}$} (1 - \tilde{\rho})^{-1} \tilde{\rho} \xi^{-1} 
g^{\alpha\beta} 
\frac{\delta \tilde{Y}_m}{\delta B^\alpha}
\frac{\delta \tilde{Y}_m}{\delta B^\beta},
\end{equation}
where the difficulty consists in determining the factor 
$(1 - \tilde{\rho})^{-1} \tilde{\rho}$ in front of the second term. Here, 
$\tilde{Y}_m$ is obtained from the functional
\begin{equation*}
Y_m = X_m + ( \hbox{$\frac{1}{2}$} \epsilon_{ab} 
\mathbf{s}_m^b \mathbf{s}_m^a + m^2 ) F 
\quad
\hbox{with}
\quad
F = \hbox{$\frac{1}{2}$} (
g_{ij} Q^i Q^j + \xi g_{\alpha\beta} \epsilon_{ab} C^{\alpha a} C^{\beta b} )
\end{equation*}
by performing the above replacements (\ref{5.1}). As is quite obvious the 
gauge fixing terms are already known from $W^{(0)}_{m, {\rm ext}}$, 
Eq.~(\ref{4.12}), so that, putting all terms together, $\tilde{Y}_m$ becomes
\begin{align}
\tilde{Y}_m &= S_{\rm cl}(A + \tilde{Q}) - H_{\alpha a} ( 
\tilde{C}^{\alpha a} - 
\tilde{\rho} \xi^{-1} g^{\alpha\beta} \epsilon^{ab} \bar{E}_{\beta b} )
\nonumber\\
&\quad + \hbox{$\frac{1}{2}$} \tilde{\sigma} g^{ij} (
\epsilon^{ab} Q^*_{i a} Q^*_{j b} - m^2 \bar{Q}_i \bar{Q}_j ) -
\tilde{\rho} \xi^{-1} g^{\alpha\beta} \epsilon^{ab} ( 
\hbox{$\frac{1}{2}$} \epsilon^{cd} E^*_{\alpha ac} E^*_{\beta bd} + 
m^2 \bar{E}_{\alpha a} \bar{E}_{\beta b} ) 
\nonumber\\
&\quad + g_{ij} \bigr( 
\tilde{Q}^i R^j_\alpha(A) \tilde{B}^\alpha +
\hbox{$\frac{1}{2}$} \epsilon_{ab}  
R^i_\alpha(A) \tilde{C}^{\alpha b} 
R^j_\beta(A + \tilde{Q}) \tilde{C}^{\beta a} + 
\hbox{$\frac{1}{2}$} m^2 \tilde{Q}^i \tilde{Q}^j \bigr)
\nonumber\\
\label{5.4}
&\quad + \xi g_{\alpha\beta} \big(
\tilde{B}^\alpha \tilde{B}^\beta - \hbox{$\frac{1}{24}$} 
\epsilon_{ab} \epsilon_{cd}  
( f^\alpha_{\gamma\rho} \tilde{C}^{\gamma a} \tilde{C}^{\rho c} )
( f^\beta_{\delta\sigma} \tilde{C}^{\delta b} \tilde{C}^{\sigma d} ) +
m^2 \epsilon_{ab} \tilde{C}^{\alpha a} \tilde{C}^{\beta b} \big) 
\\
&\quad - Q^*_{i a} R^i_\alpha(A + \tilde{Q}) \tilde{C}^{\alpha a} - 
\bar{Q}_i \bigr(
R^i_\alpha(A + \tilde{Q}) \tilde{B}^\alpha +
\hbox{$\frac{1}{2}$} \epsilon_{ab} R^i_{\alpha, j} 
R^j_\beta(A + \tilde{Q}) \tilde{C}^{\beta b} \tilde{C}^{\alpha a} \bigr)
\nonumber\\
&\quad + E^*_{\alpha ab} (
\epsilon^{ab} \tilde{B}^\alpha - \hbox{$\frac{1}{2}$}  
f^\alpha_{\beta\gamma} \tilde{C}^{\beta a} \tilde{C}^{\gamma b} ) +
\bar{E}_{\alpha a} \bigr(
f^\alpha_{\beta\gamma} \tilde{B}^\beta \tilde{C}^{\gamma a} +
\hbox{$\frac{1}{6}$} \epsilon_{cd} 
f^\alpha_{\eta\beta} f^\eta_{\gamma\delta}
\tilde{C}^{\gamma a} \tilde{C}^{\delta c} \tilde{C}^{\beta d} \bigr).
\nonumber
\end{align}
In order to facilitate a check of Eq.~(\ref{5.3}) and for later convenience 
we also write down the first and second derivative of ${\tilde Y}_m$ with 
respect to the auxiliary field $B^\alpha$:
\begin{align}
\label{5.5}
\frac{\delta_L}{\delta B^\alpha} \tilde{Y}_m +
\epsilon^{cd} E^*_{\alpha dc} &= 
2 \xi g_{\alpha\beta} \tilde{B}^\beta + 
R^j_\alpha(A) ( g_{ij} \tilde{Q}^i - \bar{Q}_j ) - 
R^i_{\alpha, j} \bar{Q}_i \tilde{Q}^j + 
f^\beta_{\alpha\gamma} \bar{E}_{\beta c} \tilde{C}^{\gamma c},
\\
\label{5.6}
\frac{\delta^2 \tilde{Y}_m}{ \delta B^\alpha \delta B^\beta} &= 
2 g_{\alpha\beta} \xi.
\end{align}
As has been shown in Appendices A and B, $S_{m, {\rm ext}}^{(0)}$ is the 
general solution of the classical master equations. By a straightforward, 
but lenghty calculation it can be verified also directly that
$S_{m, {\rm ext}}^{(0)}$ obeys the master equations (\ref{4.1}) 
subjected to the conditions (\ref{4.7}) and (\ref{4.8}). As usual, it  
serves as the tree approximation $\Gamma_m^{(0)}$ of the generating 
functional of the one--particle--irreducible vertex functions of the theory.
In addition, it may be shown that $S_{m, {\rm ext}}^{(0)}$ obeys the
properly generalized symmetries (\ref{4.14}) -- (\ref{4.16}).

After all, the fields and antifields together with the independent parameters 
of the solution (\ref{5.3}) may be multiplicatively redefined by 
corresponding $z$--factors,
\begin{gather}
z_g, \quad z_\sigma, \quad z_\rho, \quad z_\xi, \quad 
z_m^2 = z_Q^{-1} z_{\bar Q}^{-1},
\nonumber
\\
\label{5.7}
z_Q, \quad z_{\bar Q}, \quad  
z_{Q^*} = z_m^{-1} z_Q^{-1}, \quad z_F = z_m z_B^{-1},
\\
z_B, \quad z_{\bar B} = z_m^{-2} z_B^{-1}, \quad
z_{B^*} = z_m^{-1} z_B^{-1}, \quad z_C = z_m^{-1} z_B, \quad 
z_{\bar C} = z_m^{-1} z_B^{-1}, \quad z_{C^*} = z_B^{-1}. 
\nonumber
\end{gather}
Here, $z_g$ is a factor by which  
the classical action can be multiplied, 
corresponding to a redefinition of the gauge coupling constant $g$, and 
$z_\sigma$, $z_\rho$, $z_\xi$ and $z_m$ are normalization factors related
to the parameters $\sigma$, $\rho$, $\xi$ and the mass $m$, respectively.
Therefore, the general solution of the Eqs.~(\ref{4.1}) contains, in total, 
seven independent $z$--factors. Let us emphazise that this is not in 
contradiction to Ref.~\cite{16} where, reanalysing the renormalization of 
massive gauge theories due to Curci and Ferrari in Delbourgo--Jarvis gauge 
\cite{17}, only five independent $z$--factors have been found. The reason is,
that in Ref.~\cite{16} neither antiBRST-- nor $Sp(2)$--invariance has 
been required thus avoiding the $z$--factors $z_\sigma$ and $z_\rho$.

Let us now consider the $B^\alpha$--dependence of the general solution
$S_{m, {\rm ext}}^{(0)}$. From Eqs.~(\ref{5.3}) and (\ref{5.5}), (\ref{5.6}) 
it follows 
\begin{equation*}
\rho \frac{\delta_L}{\delta B^\alpha} S_{m, {\rm ext}}^{(0)} =
\frac{\delta_L}{\delta B^\alpha} \tilde{Y}_m,
\end{equation*}
which, by making use of (\ref{5.1}), may be expressed by
\begin{equation}
\label{5.8}
\rho \Bigr( \frac{\delta_L}{\delta B^\alpha} S_{m, {\rm ext}}^{(0)} + 
\epsilon^{cd} E^*_{\alpha dc} \Bigr) = 2 \xi g_{\alpha\beta} B^\beta + 
R^j_\alpha(A) ( g_{ij} Q^i - \sigma \bar{Q}_j ) - 
R^i_{\alpha, j} \bar{Q}_i Q^j + 
f^\beta_{\alpha\gamma} \bar{E}_{\beta b} C^{\gamma b}.
\end{equation}
This is just the {\it equation of motion} of the auxiliary field $B^\alpha$.
As we will show in the next Section the equations of motion for the 
(anti)ghost fields $C^{\alpha b}$ are direct consequences of Eq.~(\ref{5.8}) 
and the master equations.

Now, if the equation of motion (\ref{5.8}) is required the $z$--factors 
$z_{\bar Q}$ as well as $z_\xi$, $z_\sigma$ and $z_\rho$ are fixed 
according to
\begin{equation}
\label{5.9}
z_{\bar Q} = 1,
\qquad
z_\xi = z_Q z_B^{-1},
\qquad
z_\sigma = z_Q, 
\qquad
z_\rho = z_Q z_B,
\end{equation}
and the general solution depends only on the three independent
$z$--factors $z_g$, $z_Q$ and $z_B$, which are to be fixed by suitable
normalization conditions. 
Obviously, $\bar{Q}_i$ is not subjected to any renormalization; this 
property turns out to be essential for the study of the dependence of  the
Green's functions on the background field $A^i$ (see Section 7 below).

\smallskip
\noindent (C) {\it Generalization to higher orders}
\\
Now the solution (\ref{5.3}), including the $z$--factors, has to be 
generalized to any order of perturbation theory such that
\begin{equation*}
S_{m, {\rm ext}} = \sum_{n = 0}^\infty \hbar^n S^{(n)}_{m, {\rm ext}},
\end{equation*}
fulfills the {\it quantum} master equations (\ref{2.14}) and (\ref{2.15}) as 
well as obeys the type--I and the (broken) type--II and type--III symmetry. 
Let us emphazise that $S_{m, {\rm ext}}$ is a {\it local} functional in the 
fields and the antifields. At lowest order $S_{m, {\rm ext}}$ coincides with 
the solution $S_{m, {\rm ext}}^{(0)}$, Eq.~(\ref{5.3}). Furthermore, in order 
to ensure that $S_{m, {\rm ext}}$ depends on $\bar{B}_\alpha$ and 
$B_{\alpha b}^*$ only through the combinations (\ref{4.9}) the constraints 
(\ref{4.7}) and (\ref{4.8}) will be required to be valid at any order of 
$\hbar$; due their linearity these requirements can be realized for any 
renormalized action $S_{m, {\rm ext}}$, e.g.,~by the help of the renormalized 
quantum action principles \cite{11}. In the same manner, also the validity of 
the equation of motion for the auxiliary field $B^\alpha$ could be required.

In the following, however, we are mainly interested in the generating 
functionals of the Green's functions, $Z_m$, and the 1PI vertex functions, 
$\Gamma_m$. However, the vertex functional 
\begin{equation*}
\Gamma_{m} = \sum_{n = 0}^\infty \hbar^n \Gamma^{(n)}_{m},
\end{equation*}
is a {\em nonlocal} solution of the classical master equations (compare
Ward identities (\ref{6.3}) and (\ref{6.4}) below) which at leading order 
coincides with $S_{m, {\rm ext}}^{(0)}$.


\section{Ward identities and equations of motion}
\renewcommand{\theequation}{\thesection.\arabic{equation}}
\setcounter{equation}{0} 


Now we shall derive the Ward identities, being a consequence of the
symmetry properties of the (renormalized) extended action
$S_{m, {\rm ext}}(A|\phi, \phi_a^*, \bar{\phi}, \eta)$ which are due to 
the $osp(1,2)$--superalgebra observed by the generators $\bar{\Delta}_m^a$
and $\bar{\Delta}_A$. Moreover, in the case of background gauges,
besides of type--I invariance, additional (broken) Ward identities arise 
which are due to the type--II and type--III symmetry.

Let us introduce sources $j_I = (J_i, K_\alpha, L_{\alpha b})$ for the
fields $\phi^I = (Q^i, B^\alpha, C^{\alpha b})$. Then the extended 
generating functional of Green's functions 
$Z_m(A|j, \phi_a^*, \bar{\phi}, \eta)$ is defined as
\begin{equation}
\label{6.1}
Z_m (A|j, \phi_a^*, \bar{\phi}, \eta) = \int\, d \phi \, {\rm exp}\big\{ 
(i/\hbar) ( 
S_{m, {\rm ext}}(A|\phi, \phi_a^*, \bar{\phi}, \eta) + j_I \phi^I ) \big\}.  
\end{equation}
Multiplying Eqs.~(\ref{2.14}) and (\ref{2.15}) by 
${\rm exp}\big\{ (i/\hbar) j_I \phi^I \big\}$, integrating them over $\phi^I$,
\begin{align*}
\int\, d \phi \, {\rm exp}\big\{ (i/\hbar) j_I \phi^I \big\} \,
\bar{\Delta}_m^a \, {\rm exp}\{ (i/\hbar) S_{m, {\rm ext}} \} &= 0,
\\
\int\, d \phi \, {\rm exp}\big\{ (i/\hbar) j_I \phi^I \big\} \,
\bar{\Delta}_A \, {\rm exp}\{ (i/\hbar) S_{m, {\rm ext}} \} &= 0,
\end{align*}
and assuming, after integrating by parts, that the integrated expressions
vanish, one can rewrite the resulting equalities by the help of 
the definition (\ref{6.1}) as 
\begin{eqnarray*}
\Big\{ J_i \frac{\delta}{\delta Q^*_{i a}} + 
K_\alpha \frac{\delta}{\delta B^*_{\alpha a}} +
L_{\alpha b} \frac{\delta}{\delta C^*_{\alpha ab}} + V_m^a \Big\} Z_m = 0,
\qquad
\Big\{ (\sigma_A)_b^{~a} L_{\alpha a} \frac{\delta}{\delta F_{\alpha b}} + 
V_A \Big\} Z_m = 0,
\end{eqnarray*}
which are just the Ward identities for the generating functional of Green's
functions due to the $osp(1,2)$--symmetry of the theory.

Introducing, as usual, the 1PI vertex functional, 
$\Gamma_m(A| \phi, \phi_a^*, \bar{\phi}, \eta)$, according to
\begin{equation}
\label{6.2}
\Gamma_m = - i \hbar \,{\rm ln} Z_m - j_I \phi^I
\quad 
\hbox{with} 
\quad
\phi^I = - i \hbar \frac{\delta}{\delta j_I} {\rm ln} Z_m, 
\end{equation}
\vspace{-.5cm}
we obtain
\begin{align}
\label{6.3}
\mathbf{S}_m^a(\Gamma_m) &\equiv 
\hbox{\large $\frac{1}{2}$} ( \Gamma_m, \Gamma_m )^a - V_m^a \Gamma_m = 0,
\\
\label{6.4}
\mathbf{D}_A(\Gamma_m) &\equiv 
\hbox{\large $\frac{1}{2}$} \{ \Gamma_m, \Gamma_m \}_A - V_A \Gamma_m = 0.
\end{align}
For Yang-Mills theories the Eqs.~(\ref{6.3}) are the Slavnov-Taylor
identities of the extended BRST--symmetry. Furthermore, the Eqs.~(\ref{6.4}) 
for $A = 0$ express the ghost number conservation and, in Yang-Mills 
theories, for $A =\pm$ they are the Delduc-Sorella identities \cite{18} of the 
$Sp(2)$--symmetry.

In order to ensure type--I invariance $\Gamma_m$ will be required to fulfil 
the {\em Kluberg-Stern--Zuber identity} \cite{8} which is governed by 
\begin{gather}
\label{6.5}
\mathbf{K}_\alpha \Gamma_m = 0
\\
{\rm with} 
\qquad
\mathbf{K}_\alpha \equiv R^i_\alpha(A) \frac{\delta}{\delta A^i} + 
R^i_{\alpha, j} Q^j \frac{\delta_L}{\delta Q^i} - 
R^j_{\alpha, i} \bar{Q}_j \frac{\delta}{\delta \bar{Q}_i} + \ldots.
\nonumber
\end{gather}
Here, the ellipses $\ldots$ indicate the contributions of all the other
fields and antifields which transform homogeneously (see Eqs.~(\ref{4.17}))
and which for the following considerations are irrelevant (since they 
are identical for analogous expressions occuring below).
This identity expresses the fact that $A^i$ may be gauged arbitrarily if 
$\Gamma_m$ depends gauge--covariant on it. Let us notice, that this identity 
is defined only for $A^i \neq 0$. Hence, to fix the $A^i$--dependence of 
$\Gamma_m$ by means of an operator equation, we still need another identity
which is valid also for $A^i = 0$. Such an identity appears as a 
{\it consistency condition}:
\\
If the $B^\alpha$--dependence of $\Gamma_m$ is restricted by imposing the 
same equation of motion as it holds for ${S}^{(0)}_{m,{\rm ext}}$ then,
by virtue of the Ward identities (\ref{6.3}), that requirement leads, first 
of all, to the (anti)ghost equations of motion. Next, taking into account the 
Ward identities (\ref{6.3}) from the (anti)ghost equations of motion we 
obtain the so--called {\it Lee identity}, which is nothing else but the Ward 
identity of type--II symmetry. 

\smallskip
\noindent {\it Derivation of the Lee identity}
\\
To begin with, let us require that the $B^\alpha$-dependence of $\Gamma_m$
is governed by Eq.~(\ref{5.8}) with ${S}_{m, {\rm ext}}^{(0)}$
replaced by $\Gamma_m$,
\begin{equation}
\hspace{-.2cm}
\label{6.6}
\rho \Bigr( \frac{\delta_L}{\delta B^\alpha} \Gamma_m +
\epsilon^{cd} E^*_{\alpha dc} \Bigr) = 2 \xi g_{\alpha\beta} B^\beta + 
R^j_\alpha(A) ( g_{ij} Q^i - \sigma \bar{Q}_j ) - 
R^i_{\alpha, j} \bar{Q}_i Q^j + 
f^\beta_{\alpha\gamma} \bar{E}_{\beta b} C^{\gamma b}.
\end{equation}
Since this equation contains only terms being linear with respect to 
$Q^i$, $B^\alpha$ and $C^{\alpha b}$ its validity can be simply established 
by the help of the renormalized quantum action principles \cite{11}.
Of course, this requirement is equivalent to the related one that
the $z$--factors $z_{\bar Q}, z_\xi, z_\sigma$ and $z_\rho$, being formal
power series in $\hbar$, are determined through Eqs.~(\ref{5.9}) at any order of perturbation theory.

Let us now apply $\delta_L/\delta B^\alpha$ on the identities (\ref{6.3}) 
and (\ref{6.4}) we find as necessary conditions for $\Gamma_m$:
\begin{equation}
\label{6.7}
0 = \frac{\delta_L}{\delta B^\alpha} \mathbf{S}_m^a(\Gamma_m) =
\mathbf{\tilde Q}_m^a \frac{\delta_L}{\delta B^\alpha} \Gamma_m,
\qquad
0 = \frac{\delta_L}{\delta B^\alpha} \mathbf{D}_A(\Gamma_m) =
\mathbf{\tilde Q}_A \frac{\delta_L}{\delta B^\alpha} \Gamma_m,
\end{equation}
where the differential operators 
\begin{equation}
\label{6.8}
\mathbf{\tilde Q}_m^a X \equiv ( \Gamma_m, X )^a - V_m^a X,
\qquad
\mathbf{\tilde Q}_A X \equiv \{ \Gamma_m, X \}_A - V_A X,
\end{equation}
satisfy the $osp(1,2)$--superalgebra:
\begin{eqnarray*}
[ \mathbf{\tilde Q}_A, \mathbf{\tilde Q}_B ] = 
- \epsilon_{ABC} \mathbf{\tilde Q}^C,
\qquad
[ \mathbf{\tilde Q}_A, \mathbf{\tilde Q}_m^a ] = 
- \mathbf{\tilde Q}_m^b (\sigma_A)_b^{~a},
\qquad
\{ \mathbf{\tilde Q}_m^a, \mathbf{\tilde Q}_m^b \} = 
m^2 (\sigma^A)^{ab} \mathbf{\tilde Q}_A.
\end{eqnarray*}
Let note, that these operators are related to the untilded ones introduced
by Eqs.~(\ref{3.3}); however, they are first--order differential operators.
The first of the conditions (\ref{6.7}) can be rewritten as 
\begin{equation}
\label{6.9}
0 = \mathbf{\tilde Q}_m^a \frac{\delta_L}{\delta B^\alpha} \Gamma_m 
\equiv \mathbf{\tilde Q}_m^a \Big( 
\frac{\delta_L}{\delta B^\alpha} \Gamma_m + 
\epsilon^{cd} E^*_{\alpha dc} \Big) -
\epsilon^{ab} \Big( \frac{\delta_L}{\delta C^{\alpha b}} \Gamma_m - 
H_{\alpha b} \Big);
\end{equation}
the second condition (\ref{6.7}) gives no further constraint. Thus, 
if the $B^\alpha$--dependence of $\Gamma_m$ is fixed by imposing the 
equation of motion (\ref{6.6}) its $C^{\alpha b}$--dependence is
restricted by  
\begin{equation}
\label{6.10}
\frac{\delta_L}{\delta C^{\alpha b}} \Gamma_m - H_{\alpha b} = 
- \epsilon_{ab} \mathbf{\tilde Q}_m^a \Bigr(
\frac{\delta_L}{\delta B^\alpha} \Gamma_m + 
\epsilon^{cd} E^*_{\alpha dc} \Bigr).
\end{equation}
In fact, this is nothing else then the {\em (anti)ghost equation of motion}.
This may be shown by using the explicit expression (\ref{6.6}) for 
$\delta_L \Gamma_m/ \delta B^\alpha$:
\begin{align}
\label{6.11}
\rho \Bigr(
\frac{\delta_L}{\delta C^{\alpha b}} \Gamma_m - H_{\alpha b} \Bigr) = & 
- 2 \xi \epsilon_{ab} g_{\alpha\beta} 
\frac{\delta}{\delta B^*_{\beta a}} \Gamma_m
\\
& - R^j_\alpha(A) \Bigr(
\epsilon_{ab} g_{ij} \frac{\delta}{\delta Q^*_{i a}} \Gamma_m - 
\sigma Q^*_{j b} \Bigr) + 
\hbox{\large$\frac{1}{2}$} f^\gamma_{\alpha\beta} \epsilon_{ab} 
C^{\beta a} \frac{\delta_L}{\delta B^\gamma} \Gamma_m
\nonumber\\
& + R^i_{\alpha, j} \Bigr( 
\epsilon_{ab} \bar{Q}_i \frac{\delta}{\delta Q^*_{j a}} \Gamma_m +
Q^*_{i b} Q^j \Bigr) - f^\beta_{\alpha\gamma} \Bigr(
\epsilon_{ab} \bar{E}_{\beta c}
\frac{\delta}{\delta C^*_{\gamma ac}} \Gamma_m + 
E^*_{\beta bc} C^{\gamma c} \Bigr).   
\nonumber
\end{align}

Proceeding in the same manner as before we apply 
$\delta_L/ \delta C^{\alpha a}$ on the identities (\ref{6.3}) and 
(\ref{6.4}). This yields another set of consistency conditions:
\begin{equation}
\label{6.12}
0 = \frac{\delta_L}{\delta C^{\alpha a}} \mathbf{S}_m^a(\Gamma_m) =
- \mathbf{\tilde Q}_m^a \frac{\delta_L}{\delta C^{\alpha a}} \Gamma_m,
\qquad
0 = \frac{\delta_L}{\delta C^{\alpha a}} \mathbf{D}_A(\Gamma_m) =
\mathbf{\tilde Q}_A \frac{\delta_L}{\delta C^{\alpha a}} \Gamma_m.
\end{equation}
The first of these conditions, analogous to Eq.~(\ref{6.9}), will be 
rewritten as
\begin{equation*}
0 = - \hbox{$\frac{1}{2}$} 
\mathbf{\tilde Q}_m^a \frac{\delta_L}{\delta C^{\alpha a}} \Gamma_m \equiv 
- \hbox{$\frac{1}{2}$} \mathbf{\tilde Q}_m^a \Bigr( 
\frac{\delta_L}{\delta C^{\alpha a}} \Gamma_m - H_{\alpha a} \Bigr) + 
m^2 \Bigr(
\frac{\delta_L}{\delta B^\alpha} \Gamma_m + 
\epsilon^{cd} E^*_{\alpha dc} \Bigr),
\end{equation*}
while the other condition again gives no further constraint. From this, 
together with (\ref{6.10}), we get another identity, 
\begin{equation*}
0 = - ( \hbox{$\frac{1}{2}$} \epsilon_{ab} 
\mathbf{\tilde Q}_m^b \mathbf{\tilde Q}_m^a + m^2 ) 
\rho \Big( \frac{\delta_L}{\delta B^\alpha} \Gamma_m + 
\epsilon^{cd} E^*_{\alpha dc} \Big) \equiv 
\mathbf{L}_\alpha \Gamma_m - m^2 g_{ij} R^i_\alpha(A) Q^j.
\end{equation*}
Inserting for $\delta_L \Gamma_m/\delta B^\alpha$ the expression (\ref{6.6}) 
we obtain the explicit form of the Lee identity:
\begin{gather}
\label{6.13}
\mathbf{L}_\alpha \Gamma_m = m^2 g_{ij} R^i_\alpha(A) Q^j
\\
\hbox{with}
\quad
\mathbf{L}_\alpha \equiv \sigma R^i_\alpha(A + \sigma^{-1} Q)  
\frac{\delta_L}{\delta Q^i} + 
g_{ij} R^i_\alpha(A + \bar{Q}) \frac{\delta}{\delta \bar{Q}_j} + \ldots,
\nonumber
\end{gather}
where $R^i_\alpha(Q) \equiv R^i_\alpha(0) + R^i_{\alpha, j} Q^j$ and
$R^i_\alpha(\bar{Q}) \equiv R^i_\alpha(0) - g^{ij} R^k_{\alpha, j} \bar{Q}_k$.
It can be checked that the Lee operator $\mathbf{L}_\alpha$ is 
independent of  $\rho$ and $\xi$. 
The term $m^2 g_{ij} R^i_\alpha(A) Q^j$ stems from the
fact that the type--II invariance is broken by that mass term 
(compare Eq.~(\ref{4.15})).

In writing down the identity (\ref{6.13}) one still has to require that 
$\Gamma_m$ obeys constraints implied by the equations: 
\begin{gather*}
\hbox{\large $\frac{1}{2}$} \epsilon_{cd}
\frac{\delta}{\delta C^*_{\alpha dc}} \Gamma_m = B^\alpha,
\qquad
\frac{\delta}{\delta \bar{B}_\alpha} \Gamma_m =  m^2 B^\alpha,
\\
\frac{\delta}{\delta F_{\alpha a}} \Gamma_m = - C^{\alpha a},
\qquad
\frac{\delta}{\delta B^*_{\alpha a}} \Gamma_m + 
\hbox{\large $\frac{1}{2}$} 
\frac{\delta}{\delta \bar{C}_{\alpha a}} \Gamma_m = m^2 C^{\alpha a},
\end{gather*}
which generalize the corresponding requirements for $\Gamma^{(0)}_{m}$
to any orders. Again, their validity can be simply established by using
 the quantum action principles \cite{11}.


\section{Background dependence of Green's functions}
\renewcommand{\theequation}{\thesection.\arabic{equation}}
\setcounter{equation}{0} 


Now, having completely characterized the symmetry properties of $\Gamma_m$ 
let us enquire into its $A^i$--dependence. This is achieved by comparing the
Kluberg-Stern--Zuber identity (\ref{6.5}) with the Lee identity (\ref{6.13}) 
which leads to the differential equation
\begin{equation}
\label{7.1}
\frac{\delta}{\delta A^i} \Gamma_m = \Bigr(
\sigma \frac{\delta_L}{\delta Q^i} + 
g_{ij} \frac{\delta}{\delta \bar{Q}_j} \Bigr) \Gamma_m - m^2 g_{ij} Q^j
\end{equation}
describing the $A^i$-dependence of 
$\Gamma_m(A| \phi, \phi_a^*, \bar{\phi}, \eta)$. Its integration yields
\begin{align}
\label{7.2}
\Gamma_m(A| \phi, \phi_a^*, \bar{\phi}, \eta) = \,& 
{\rm exp}\Bigr\{ A^i \Bigr( 
\sigma \frac{\delta_L}{\delta Q^i} + 
g_{ij} \frac{\delta}{\delta \bar{Q}_j} \Bigr) \Bigr\} 
\Gamma_m(0| \phi, \phi_a^*, \bar{\phi}, \eta)
\\
& - m^2 g_{ij} \bigr( A^i Q^j + \hbox{$\frac{1}{2}$} \sigma A^i A^j \bigr),
\nonumber
\end{align}
i.e.,~the $A^i$--dependence is completely determined by the $Q^i$-- and 
$\bar{Q}_i$--dependence of the corresponding functional 
$\Gamma_m(0| \phi, \phi_a^*, \bar{\phi}, \eta)$ for $A^i = 0$.

By making use of Eq.~(\ref{7.1}) the Lee identity (\ref{6.13}) can be cast 
into the form
\begin{gather}
\label{7.3}
\mathbf{W}_\alpha \Gamma_m = m^2 g_{ij} R^i_\alpha(0) Q^j
\\
\hbox{with} 
\qquad
\mathbf{W}_\alpha \equiv  
\sigma R^i_\alpha(\sigma^{-1} Q) \frac{\delta_L}{\delta Q^i} +
g_{ij} R^j_\alpha(\bar{Q}) \frac{\delta}{\delta \bar{Q}_i} + \ldots,
\nonumber
\end{gather}
which is just the Ward identity of type--III symmetry. Integrating 
(\ref{7.3}) over space-time we recover the Ward identity of the {\it rigid} 
symmetry,
\begin{gather}
\label{7.4}
\int d^4x\, \mathbf{R}_\alpha \Gamma_m = 0
\\
\hbox{with}
\qquad
\mathbf{R}_\alpha \equiv  
R^i_{\alpha, j} A^j \frac{\delta}{\delta A^i} + 
R^i_{\alpha, j} Q^j \frac{\delta_L}{\delta Q^i} -
R^j_{\alpha, i} \bar{Q}_j \frac{\delta}{\delta \bar{Q}_i} + \ldots.
\nonumber
\end{gather}
For the relation between $Z_m(A| j, \phi_a^*, \bar{\phi}, \eta)$ and 
$Z_m(0| j, \phi_a^*, \bar{\phi}, \eta)$ corresponding to Eq.~(\ref{7.1}),
by virtue of (\ref{6.2}), we obtain the differential equation 
\begin{equation*}
\frac{\delta}{\delta A^i} Z_m = g_{ij} \Bigr(
\frac{\delta}{\delta \bar{Q}_j} - m^2 \frac{\delta}{\delta J_j} \Bigr) Z_m - 
(i/\hbar) \sigma J_i Z_m.
\end{equation*}
Its solution is simply given by 
\begin{align}
\label{7.5}
Z_m(A|j, \phi_a^*, \bar{\phi}, \eta) = \,&{\rm exp}
\Bigr\{\! - (i/\hbar) \sigma \bigr( 
J_i A^i - \hbox{$\frac{1}{2}$} m^2 g_{ij} A^i A^j \bigr) \Bigr\} \times 
\\
& {\rm exp}\Bigr\{ g_{ij} A^i \Bigr(
\frac{\delta}{\delta \bar{Q}_j} - m^2 \frac{\delta}{\delta J_j} \Bigr) \Bigr\} 
Z_m(0| j, \phi_a^*, \bar{\phi}, \eta).
\nonumber
\end{align}
Together with Eq.~(\ref{7.2}) this is the main result of this Section.

\smallskip
\noindent {\em Background dependence of physical Green's functions}\\
In order to apply this exact relation we consider {\it physical} 
Green's functions, i.e.,~on--shall Green's functions
$Z_{\rm phys}(A|J) := Z\big(A|J, \{ K, L_b; \phi_a^*, \bar{\phi}, \eta \} = 0\big)$ 
of transverse gauge fields,
$g_{ij} R^i_\alpha(A) (\delta/\delta J_j) Z_{\rm phys}(A|J) = 0$,
for $m = 0$ and choosing the Landau gauge $\xi = 0$.
In doing that, we first set $A^i = 0$ and put -- with the exception of 
$J_i$ and $\bar{Q}_i$ -- any of the sources in 
$Z_m(0| j, \phi_a^*, \bar{\phi}, \eta)$ equal to zero. Then, the problem 
consists in determining the explicit $\bar{Q}_i$--dependence of 
$Z_m(0|J, \bar{Q})$.
 
To begin with, let us write down the rigid Ward identity (\ref{7.4}) and 
the Lee identity (\ref{7.3}) for the particular case under consideration,
\begin{alignat}{2}
\label{7.6}
\int d^4x\, \mathbf{R}_\alpha Z_m(0|J, \bar{Q}) &= 0,
&\qquad
\mathbf{R}_\alpha &= - R^i_{\alpha, j} \Bigr(
J_i \frac{\delta}{\delta J_j} + 
\bar{Q}_i \frac{\delta}{\delta \bar{Q}_j} \Bigr),
\\
\Bigr( \mathbf{L}_\alpha - m^2 g_{ij} R^i_\alpha(0)
\frac{\delta}{\delta J_j} \Bigr) Z_m(0|J, \bar{Q}) &= 0,
&\qquad
\mathbf{L}_\alpha &= R^i_\alpha(0) \Bigr( 
- (i/\hbar) \sigma J_i + 
g_{ij} \frac{\delta}{\delta \bar{Q}_j} \Bigr) + \mathbf{R}_\alpha,
\nonumber
\end{alignat}
where the operator $\mathbf{L}_\alpha$ differs from $\mathbf{R}_\alpha$ only
by divergence terms, i.e.,~terms proportional to $R^i_\alpha(0)$ which
vanish after integration over space--time. 

In the limit $\bar{Q}_i = 0$ the following general ansatz for
$\mathbf{L}_\alpha$ has to be taken when applied to $Z_m(0|J,0)$:
\begin{alignat}{2}
\label{7.7}
\int d^4x\, \mathbf{R}_\alpha Z_m(0|J,0) &= 0,
&\qquad
\mathbf{R}_\alpha &= - R^i_{\alpha, j} J_i \frac{\delta}{\delta J_j},  
\\
\Big( \mathbf{L}_\alpha - m^2 g_{ij} R^i_\alpha(0)
\frac{\delta}{\delta J_j} \Big) Z_m(0|J,0) &= 0,
&\qquad
\mathbf{L}_\alpha &= R^i_\alpha(0) \bigr( - (i/\hbar) z_Q^{-1} J_i \bigr) +
\mathbf{R}_\alpha,
\nonumber
\end{alignat}
with $z_Q^{-1}$ being the $z$--factor of $J_i$. The necessity for
making that ansatz results from the fact that in the case $\bar{Q}_i=0$
the identities (\ref{6.13}) and (\ref{7.3}) are undefined.
Comparing (\ref{7.6}) and (\ref{7.7}) the $\bar{Q}_i$--dependence of 
$Z_m(0|J, \bar{Q})$ is obtained:
\begin{equation}
\label{7.8}
Z_m(0|J, \bar{Q}) = {\rm exp}\Bigr\{  
(i/\hbar) (\sigma - z_Q^{-1}) g^{ij} \bigr(
J_i \bar{Q}_j + \hbox{$\frac{1}{2}$} m^2 \bar{Q}_i \bar{Q}_j \bigr) \Bigr\} 
\,{\rm exp}\Bigr\{ m^2 \bar{Q}_i \frac{\delta}{\delta J_i} \Bigr\} Z_m(0|J).
\end{equation}
This relation immediately leads to the conclusion that, in the absence of 
sources $J_i$ and $\bar{Q}_i$, the following equality holds 
$\delta Z(0)/ \delta \bar{Q}_i \equiv \bigr(
\delta Z_m(0| J, \bar{Q})/ \delta \bar{Q}_i \bigr)
|_{J = \bar{Q} = m = 0} = 0$. 
This property, which for the analogous case of the background field $A^i$ 
instead of $\bar{Q}_i$ already has been noted  in the first of 
Refs.~\cite{8}, turns out to be essential for the study of the 
$A^i$--dependence.

Now, we are able to consider the case $A^i \neq 0$. Inserting
the expression (\ref{7.8}) for $Z_m(0|J, \bar{Q})$ into the relation
(see Eq.~(\ref{7.5}))
\begin{equation*}
Z_m(A|J, \bar{Q}) = {\rm exp}\Bigr\{\! - \frac{i}{\hbar} \sigma \bigr( 
J_i A^i - \hbox{$\frac{1}{2}$} m^2 g_{ij} A^i A^j \bigr) \Big\}
\, {\rm exp}\Bigr\{ g_{ij} A^i \Bigr(
\frac{\delta}{\delta \bar{Q}_j} - m^2 \frac{\delta}{\delta J_j} \Bigr) \Bigr\} 
Z_m(0|J, \bar{Q}),
\end{equation*}
we get 
\begin{align*}
Z_m(A|J, \bar{Q}) = \,& {\rm exp}\Bigr\{ (i/\hbar) 
(\sigma - z_Q^{-1}) g^{ij} \bigr( J_i \bar{Q}_j +
\hbox{$\frac{1}{2}$} m^2 \bar{Q}_i \bar{Q}_j \bigr) \Bigr\} \times
\\
& {\rm exp}\Big\{\! - (i/\hbar) z_Q^{-1} \bigr( J_i A^i -
\hbox{$\frac{1}{2}$} m^2 g_{ij} A^i A^j \bigr) \Bigr\} 
\,{\rm exp}\Bigr\{\! - m^2 g_{ij} A^i \frac{\delta}{\delta J_j} \Bigr\} 
Z_m(0|J,0).
\end{align*}
Here, we can put $\bar{Q}_i = 0$ and obtain the relation between 
$Z_m(A|J)$ and $Z_m(0|J)$ we are looking for, namely
\begin{equation}
\label{7.9}
Z_m(A|J) = {\rm exp}\Bigr\{\! - (i/\hbar) z_Q^{-1} \bigr( J_i A^i - 
\hbox{$\frac{1}{2}$} m^2 g_{ij} A^i A^j \bigr) \Bigr\}
\, {\rm exp}\Bigr\{ - m^2 g_{ij} A^i \frac{\delta}{\delta J_j} \Bigr\} 
Z_m(0|J).
\end{equation}
Notice, that for $m = 0$ and choosing the Landau gauge $\xi = 0$ this 
relation coincides with a corresponding result obtained by Rouet using 
a quite different method \cite{14}:
\begin{equation*}
Z_{\rm phys}(A|J) = {\rm exp}\bigr\{ - (i/\hbar) z_Q^{-1} J_i A^i \bigr\}
Z_{\rm phys}(0|J).
\end{equation*}
This relation states that physical Green's functions in a background gauge
are obtained from the Green's functions without background field by a mere
translation.

Finally, let us point at the similarity between the relations 
(\ref{7.8}) and (\ref{7.9}), which indicates that ${\bar Q}_i$ for 
$A^i = 0$ plays the role of a background field (because of the above 
mentioned property $\delta Z(0)/ \delta \bar{Q}_i = 0$ in the absence of $J_i$ 
and $\bar{Q}_i$). That is the reason why generally one should require not 
only background gauge (type--I) invariance but also quantum gauge 
(type--II and type--III) invariance: 
both symmetry requirements together ensure that neither $A^i$ nor $\bar{Q}_i$ 
are subjected to any renormalization.


\section{Ultraviolet asymptotics}
\renewcommand{\theequation}{\thesection.\arabic{equation}}
\setcounter{equation}{0} 


Now we are going to show that the ultraviolet asymptotics of $\Gamma_m$
is independent of the background field $A^i$. In order to obtain this result
we derive the renormalization group equation and prove, making use of the
basic relation (\ref{7.3}), that the $\beta$--function and the anomalous 
dimensions in the presence of a background field $A^i$ agree with 
the corresponding ones for $A^i = 0$.

For a consistent treatment of the ultraviolet divergences emerging from the
Feynman graphs the BPHZL subtraction scheme \cite{19} will be employed 
which allows the application of  the renormalized quantum action principles 
\cite{11}. In this scheme the mass $m$ is replaced by $m_s = (1 - s) m$,
where the parameter $s$ ($0 \leq s \leq 1$) is incorporated in order to
avoid spurious infrared singularities which would occure by ultraviolet 
subtractions at $s = 1$; it interpolates between the massive ($s = 0$) and 
the massless ($s = 1$) case.  

In order to fix $\Gamma_m$ completely we have to choose suitable
normalization conditions for the independent $z$-factors $z_g$, $z_Q$
and $z_B$, being formal power series in $\hbar$ (not to be specified further). 
Denoting the  (Euclidean) normalization point by $\mu^2$, the physical content 
of the theory must be independent on how $\mu^2$ is chosen. This requirement 
is expressed by the renormalization group equation which relates the action of
$\nabla \equiv \mu^2 \partial/ \partial \mu^2$ on $\Gamma_m$
to a corresponding change of the independent parameters of the theory.

As is well known $\nabla\Gamma_m$, by the quantum action principle, defines 
an insertion $\Delta$:
\begin{equation}
\label{8.1}
\nabla \Gamma_m = \Delta \cdot \Gamma_m = \Delta + O(\hbar \Delta).
\end{equation}
Now, it is our task to expand this insertion $\Delta$ into a suitable basis 
of independent symmetric insertions. In general, an insertion $\Delta$ is 
called {\it symmetric} if the corresponding differential operator $\nabla$ 
satisfies a set of constraints which are related to the symmetry properties 
of $\Gamma_m$:

First, as  $\Gamma_m$ is assumed to fulfil the Ward identities of both the
(anti)BRST-- and $Sp(2)$--symmetry,
\begin{equation*}
\mathbf{S}_m^a(\Gamma_m) = 0,
\qquad
\mathbf{D}_A(\Gamma_m) = 0,
\end{equation*}
the operator $\nabla$ is restricted to 
\begin{equation}
\label{8.2}
[ \mathbf{\tilde Q}_m^a, \nabla ] \Gamma_m = 0,
\qquad
[ \mathbf{\tilde Q}_A, \nabla ] \Gamma_m = 0,
\end{equation}
where
$\mathbf{\tilde Q}_m^a$ and $\mathbf{\tilde Q}_A$ are defined by
Eqs.~(\ref{6.8}). Second, because $\Gamma_m$ obeys both the equations of 
motion for the auxiliary fields (see Eq.~(\ref{6.6})),
\begin{align*}
\mathbf{B}_\alpha \Gamma_m &= - \rho \epsilon^{cd} E^*_{\alpha dc} + 
2 \xi g_{\alpha\beta} B^\beta + 
R^j_\alpha(A) ( g_{ij} Q^i - \sigma \bar{Q}_j ) - 
R^i_{\alpha, j} \bar{Q}_i Q^j + 
f^\beta_{\alpha\gamma} \bar{E}_{\beta b} C^{\gamma b},
\\
\mathbf{B}_\alpha &= \rho \frac{\delta}{\delta B^\alpha},
\\
\intertext{and the (anti)ghost fields (see Eq.~(\ref{6.11})):}
\mathbf{C}_{\alpha b} \Gamma_m &= \rho H_{\alpha b} - 
\sigma R^j_\alpha(A) Q^*_{j b} +  
R^i_{\alpha, j} Q^*_{i b} Q^j - 
f^\beta_{\alpha\gamma} E^*_{\beta bc} C^{\gamma c},   
\\
\mathbf{C}_{\alpha b} &= 
\rho \frac{\delta}{\delta C^{\alpha b}} + 
2 \xi \epsilon_{ab} g_{\alpha\beta} \frac{\delta}{\delta B^*_{\beta a}}
\\
&\quad + \epsilon_{ab} \bigr(
g_{ij} R^j_\alpha(A) - R^i_{\alpha, j} \bar{Q}_i \bigr) 
\frac{\delta}{\delta Q^*_{i a}} - 
\hbox{\large $\frac{1}{2}$} f^\gamma_{\alpha\beta} \epsilon_{ab} 
C^{\beta a} \frac{\delta}{\delta B^\gamma} +
f^\beta_{\alpha\gamma} \epsilon_{ab} \bar{E}_{\beta c} 
\frac{\delta}{\delta C^*_{\gamma ac}},
\end{align*}
$\nabla$ will be required further to satisfy the following relations:
\begin{equation}
\label{8.3}
[ \mathbf{B}_\alpha, \nabla ] \Gamma_m = 0,
\qquad
[ \mathbf{C}_{\alpha b}, \nabla ] \Gamma_m = 0;
\end{equation}
finally, since $\Gamma_m$ obeys the Ward identities of type--I, type--II 
and type--III invariance
(see Eqs.~(\ref{6.5}), (\ref{6.13}) and (\ref{7.3})), respectively,
\begin{equation*}
\mathbf{K}_\alpha \Gamma_m = 0,
\qquad
\mathbf{L}_\alpha \Gamma_m = m^2 g_{ij} R^i_\alpha(A) Q^j,
\qquad
\mathbf{W}_\alpha \Gamma_m = m^2 g_{ij} R^i_\alpha(0) Q^j,
\end{equation*}
$\nabla$ should also satisfy the following relations:
\begin{equation}
\label{8.4}
[ \mathbf{K}_\alpha, \nabla ] \Gamma_m = 0,
\qquad
[ \mathbf{L}_\alpha, \nabla ] \Gamma_m = 0,
\qquad
[ \mathbf{W}_\alpha, \nabla ] \Gamma_m = 0.
\end{equation}

Since neither $\mathbf{\tilde Q}_m^a$ and $\mathbf{\tilde Q}_A$ nor
$\mathbf{B}_\alpha$, $\mathbf{C}_{\alpha b}$, $\mathbf{K}_\alpha$,
$\mathbf{L}_\alpha$ and $\mathbf{W}_\alpha$ depend on the normalization point 
$\mu$, the operator $\nabla$ obiously is {\it symmetric}. Hence, our task is 
to construct a {\it basis} for any symmetric differential operator in a form
which permits a generalization to all orders of $\hbar$. This is most easily 
done by using the results of Section 5; actually one needs only the 
Eqs.~(\ref{5.7}) and (\ref{5.9}) for the $z$-factors.    

A basis of symmetric operators which fulfil the requirements (\ref{8.2}), 
as can be deduced from Eqs.~(\ref{5.7}), consists of seven differential 
operators which are given by
\begin{equation*} 
g \frac{\partial}{\partial g}, 
\qquad
\xi \frac{\partial}{\partial \xi}, 
\qquad
\sigma \frac{\partial}{\partial \sigma}, 
\qquad
\rho \frac{\partial}{\partial \rho}
\end{equation*} 
and the following counting operators:
\begin{align*}
\mathbf{N}_Q &= \bar{B}_\alpha \frac{\delta }{\delta \bar{B}_\alpha} +
Q^i \frac{\delta_L}{\delta Q^i} + \hbox{\large$\frac{1}{2}$} \Bigr(
B^*_{\alpha a} \frac{\delta }{\delta B^*_{\alpha a}} -
Q^*_{i a} \frac{\delta }{\delta Q^*_{i a}} \Bigr)
\\
&\quad + \hbox{\large$\frac{1}{2}$} \Bigr(
C^{\alpha b} \frac{\delta_L}{\delta C^{\alpha b}} +
\bar{C}_{\alpha b} \frac{\delta }{\delta \bar{C}_{\alpha b}} -
F_{\alpha b} \frac{\delta }{\delta F_{\alpha b}} \Bigr) -
m^2 \frac{\partial }{\partial m^2},
\\
\mathbf{N}_{\bar Q} &= \bar{B}_\alpha \frac{\delta }{\delta \bar{B}_\alpha} +
\bar{Q}_i \frac{\delta }{\delta \bar{Q}_i} +
\hbox{\large$\frac{1}{2}$} \Bigr(
B^*_{\alpha a} \frac{\delta }{\delta B^*_{\alpha a}} +
Q^*_{i a} \frac{\delta }{\delta Q^*_{i a}} \Bigr)
\\
&\quad + \hbox{\large$\frac{1}{2}$} \Bigr(
C^{\alpha b} \frac{\delta_L}{\delta C^{\alpha b}} +
\bar{C}_{\alpha b} \frac{\delta }{\delta \bar{C}_{\alpha b}} -
F_{\alpha b} \frac{\delta }{\delta F_{\alpha b}} \Bigr) -
m^2 \frac{\partial }{\partial m^2},
\\
\mathbf{N}_B &= B^\alpha \frac{\delta_L }{\delta B^\alpha} -
\bar{B}_\alpha \frac{\delta }{\delta \bar{B}_\alpha} -
B^*_{\alpha a} \frac{\delta }{\delta B^*_{\alpha a}}
\\
&\quad + C^{\alpha b} \frac{\delta_L}{\delta C^{\alpha b}} -
\bar{C}_{\alpha b} \frac{\delta }{\delta \bar{C}_{\alpha b}} -
C^*_{\alpha ab} \frac{\delta }{\delta C^*_{\alpha ab}} -
F_{\alpha b} \frac{\delta }{\delta F_{\alpha b}}.
\end{align*}
Hence, any $osp(1,2)$--symmetric differential operator $\nabla$ can be 
expanded with respect to this basis:
\begin{equation}
\label{8.5}
\nabla = \beta_g g \frac{\partial}{\partial g} +
\beta_\xi \xi \frac{\partial}{\partial \xi} +
\beta_\sigma \sigma \frac{\partial}{\partial \sigma} +
\beta_\rho \rho \frac{\partial}{\partial \rho} +
\gamma_Q \mathbf{N}_Q + 
\gamma_{\bar Q} \mathbf{N}_{\bar Q} + 
\gamma_B \mathbf{N}_B.  
\end{equation}
Obviously, these symmetric differential operators are independent of the
background field.
 
Now, requiring that $\nabla$ satisfies also the constraints (\ref{8.3})
and (\ref{8.4}) leads, in accordance with (\ref{5.9}), to the restrictions 
\begin{gather*}
\gamma_{\bar Q} = 0,
\qquad
\beta_\xi = \gamma_Q - \gamma_B,
\qquad
\beta_\sigma = \gamma_Q,
\qquad
\beta_\rho = \gamma_Q + \gamma_B,
\end{gather*}
so that the expansion (\ref{8.5}) reduces to
\begin{equation}
\label{8.6}
\nabla = \beta_g g \frac{\partial}{\partial g} +
\gamma_Q \mathbf{\tilde{N}}_Q + \gamma_B \mathbf{\tilde{N}}_B,  
\end{equation}
where $\mathbf{\tilde{N}}_Q$ and $\mathbf{\tilde{N}}_B$ are given by
\begin{equation*}
\mathbf{\tilde{N}}_Q \equiv \mathbf{N}_Q + 
\xi \frac{\partial}{\partial \xi} +
\sigma \frac{\partial}{\partial \sigma} +
\rho \frac{\partial}{\partial \rho},
\qquad
\mathbf{\tilde{N}}_B \equiv \mathbf{N}_B - 
\xi \frac{\partial}{\partial \xi} +
\rho \frac{\partial}{\partial \rho}.
\end{equation*}

Now, having defined to any order of $\hbar$ a basis of symmetric differential 
operators let us return to our starting point: Expanding 
$\Delta \cdot \Gamma_m$ on the right--hand side of Eq.~(\ref{8.1}) 
on the basis of symmetric insertions generated by the operator (\ref{8.6}) 
the renormalization group equation in the background gauge reads
\begin{equation}
\label{8.7}
\mu^2 \frac{\partial}{\partial \mu^2} 
\Gamma_m(A| \phi, \phi_a^*, \bar{\phi}, \eta) = 
\bigr( \beta_g g \frac{\partial}{\partial g} +
\gamma_Q \mathbf{\tilde{N}}_Q + \gamma_B \mathbf{\tilde{N}}_B \bigr) 
\Gamma_m(A| \phi, \phi_a^*, \bar{\phi}, \eta),  
\end{equation}
where the $\beta$--function $\beta_g$ and the anomalous dimensions $\gamma_Q$, 
$\gamma_B$ start at first order in $\hbar$. Next, by virtue of 
Eq.~(\ref{7.2}), from the previous relation (\ref{8.7}) one obtains
\begin{equation}
\label{8.8}
\mu^2 \frac{\partial}{\partial \mu^2} 
\Gamma_m(0| \phi, \phi_a^*, \bar{\phi}, \eta) = 
\bigr( \beta_g g \frac{\partial}{\partial g} +
\gamma_Q \mathbf{\tilde{N}}_Q + \gamma_B \mathbf{\tilde{N}}_B \bigr) 
\Gamma_m(0| \phi, \phi_a^*, \bar{\phi}, \eta),  
\end{equation}
which is just the renormalization group equation for $A^i = 0$. Obviously, 
the coefficients $\beta_g$, $\gamma_Q$ and $\gamma_B$ in both renormalization 
group equations are the same. In addition, it should be emphasized that by 
choosing suitable normalization conditions for $A^i = 0$ the solution of the 
renormalization group equation (\ref{8.7}) is uniquely determined by the 
corresponding solution of the renormalization group equation (\ref{8.8})
because by virtue of Eq.~(\ref{7.2}) the normalization conditions for
$A^i \neq 0$ are fixed as well. Thus, it is proven that the ultraviolet 
asymptotics of the vertex functions is in fact independent of the 
background field $A^i$ to all orders of perturbation theory.


\section{Concluding remarks}
\renewcommand{\theequation}{\thesection.\arabic{equation}}
\setcounter{equation}{0} 


Let us first state the essential results which have been obtained.
Under the assumption of a {\it linear} quantum--background splitting it has 
been shown that for irreducible massive gauge theories of first rank
with a generic background field the renormalized generating functional 
$\Gamma_m(A)$ is invariant under background (type--I) as well as quantum 
(type--II and type--III) gauge transformations, thereby generalizing the 
results of Ref.~\cite{8} by exploiting the full non--minimal sector of
the theory. As a consequence of this we were able to determine the 
$A^i$--dependence of $\Gamma_m(A)$ completely by these symmetry requirements.
In order to determine the $A^i$--dependence of $\Gamma_m(A)$ explicitely 
one only has to determine the dependence of $\Gamma_m(0)$ on the gauge 
fields $Q^i$ and the associated antifields $\bar{Q}_i$. As an application
we independently recovered an earlier result of Rouet \cite{14}. Furthermore, 
it was proven that introducing a background gauge does not change the 
ultraviolet asymptotics of the theory. 

In this paper only generic background configurations, i.e.,~the gauge
zero modes, are taken into account. However, within the quite general frame
we introduced here our results may be generalized.
If the classical action is not only gauge invariant but has additional 
symmetries and if $A^i$ is a solution of the equation of motion for 
$Q^i$ depending on collective coordinates which break the additional
symmetries, then similar conclusions can be drawn.
In that case, according to the method of Gervais and Sakita \cite{20},
one first has to factor out the dependence of the generating functional 
$Z_m(A)$ on the collective coordinates of $A^i$. After that, our method
can be applied in order to reduce the determination of the background
dependence of $\Gamma_m(A)$ to quantities being independent of $A^i$, i.e.,
to the determination of the antifield dependence of $\Gamma_m(0)$. If
the antifield dependence of $\Gamma_m$ has been determined once and for all
the result is valid for any background configuration having the same 
dependence on the collective coordinates.

In the case of a {\it nonlinear} quantum--background splitting  -- which takes
place, e.g., for $N = 1$ supersymmetric theories -- the situation is much more 
involved, since then both the type--II and type--III symmetries are 
{\it nonlinear} ones. In this case the determination of the background 
dependence turns out to be very complicated.


\bigskip\bigskip
\noindent
{\bf \large Acknowldgement}
\bigskip

\noindent
The authors like to thank P. Lavrov for many discussion concerning
the general aspects of the $Sp(2)$--quantization procedure. They
also like to thank the referee for critical remarks which   
led to an improvement of our arguments whose hard core has been
given in the two Appendixes.

\bigskip

\begin{appendix}


\section{Construction of the general solution  of the classical master 
equations (\ref{4.11}) with vanishing new ghost number}
\renewcommand{\theequation}{\thesection.\arabic{equation}}
\setcounter{equation}{0} 


In this Appendix, we show that the expression (\ref{4.11}) is the general 
solution of the classical master equations (\ref{4.1}), with vanishing 
new ghost number, ${\rm ngh}(W_m^{(0)}) = 0$, i.e.~being linear in the 
antifields,  
\begin{equation}
\label{A0}
\hbox{\large $\frac{1}{2}$} ( W^{(0)}_m, W^{(0)}_m)^a - 
V_m^a W^{(0)}_m = 0,
\qquad
\hbox{\large $\frac{1}{2}$} \{ W^{(0)}_m, W^{(0)}_m \}_A - 
V_A W^{(0)}_m = 0, 
\end{equation}
and being
subjected to the constraints (\ref{4.7}), (\ref{4.8}). For that reason we 
ascribe according to \cite{2} to all quantities, including the background 
field $A^i$ and formally also the mass parameter $m$, the following new ghost 
numbers: 
\begin{alignat*}{2}
{\rm ngh}(Q^i, C^{\alpha b}, B^\alpha) &= (0, 1, 2),
&\qquad 
{\rm ngh}(A^i) &= 0, 
\\
{\rm ngh}(Q_{i a}^*, C_{\alpha ab}^*, B_{\alpha a}^*) &= (-1, -2, -3), 
&\qquad 
{\rm ngh}(m) &= 1, 
\\
{\rm ngh}(\bar{Q}_i, \bar{C}_{\alpha b}, \bar{B}_\alpha) &= (-2, -3, -4),
&\qquad 
{\rm ngh}(F_{\alpha b}) &= -1. 
\end{alignat*}

Now, adopting the method of Ref.~\cite{2} the solution $W_m^{(0)}$ will be 
sought in the form of an terminating expansion
\begin{equation}
\label{A.1}
W_m^{(0)} = S_{\rm cl}(A + Q) + \sum_{n = 1} W_{m(n)},
\qquad
{\rm ngh}(W_{m(n)}) = 0,
\end{equation}
where $W_{m(n)}$ are polynomials of $n$th order in powers of the fields
$B^\alpha$ and $C^{\alpha a}$. Furthermore, $W_{m(n)}$ is subjected to the 
above mentioned constraints (\ref{4.7}) and (\ref{4.8}) which now read 
\begin{equation*}
\hbox{\large$\frac{1}{2}$} \epsilon_{ab}
\frac{\delta}{\delta C^*_{\alpha ba}} W_{m(n)} = B^\alpha,
\qquad
\frac{\delta}{\delta F_{\alpha a}} W_{m(n)} = - C^{\alpha a},
\end{equation*}
\begin{equation*}
\bigr( \frac{\delta}{\delta B^*_{\alpha a}} +
\hbox{\large$\frac{1}{2}$} \frac{\delta}{\delta \bar{C}_{\alpha a}} \bigr) 
W_{m(n)} = - m^2 \frac{\delta}{\delta F_{\alpha a}} W_{m(n)},
\qquad
\frac{\delta}{\delta \bar{B}_\alpha} W_{m(n)} = 
\hbox{\large$\frac{1}{2}$} m^2 \epsilon_{ab}
\frac{\delta}{\delta C^*_{\alpha ba}} W_{m(n)};
\end{equation*}
thus, the polynomials $W_{m(n)}$ depend on $B^*_{\alpha a}$ and 
$\bar{B}_\alpha$ only through the combinations (\ref{4.9}),
\begin{equation*}
\bar{E}_{\alpha a} = \bar{C}_{\alpha a} - 
\hbox{$\frac{1}{2}$} B^*_{\alpha a},
\qquad 
E^*_{\alpha ab} = C^*_{\alpha ab} - 
\hbox{$\frac{1}{2}$} m^2 \epsilon_{ab} \bar{B}_\alpha,
\qquad
H_{\alpha a} = F_{\alpha a} - m^2 B^*_{\alpha a}.
\end{equation*}

Let us consider the first approximation $W_{m(1)}$. The most general form 
of $W_{m(1)}$ which fulfils the above mentioned requirements is
\begin{equation*}
W_{m(1)} = - ( \bar{Q}_i \Lambda^i_\alpha - 
\epsilon^{ab} E_{\alpha ab}^* ) B^\alpha - 
( Q_{i b}^* \Lambda^{i b}_{\alpha a} + H_{\alpha a} ) C^{\alpha a} +
\hbox{$\frac{1}{2}$} Q_{i a}^* Q_{j b}^* \Lambda^{ia jb}_\alpha B^\alpha,
\end{equation*}
where $\Lambda^i_\alpha$, $\Lambda^{i b}_{\alpha a}$ and
$\Lambda^{ia jb}_\alpha$ are some unknown matrices depending on the
fields $A^i$ and $Q^i$. Next, we require that $S_{\rm cl}(A + Q) + W_{m(1)}$ 
fulfils the classical master equations to first order. 
This leads to the result that $\Lambda^i_\alpha = R^i_\alpha(A + Q)$ and
$\Lambda^{i b}_{\alpha a} = R^i_\alpha(A + Q) \delta^b_a$ has to be identified
with the gauge generators and that $\Lambda^{ia jb}_\alpha = 0$ has to be put
equal to zero. Thus, the first approximation is given by
\begin{equation}
\label{A.2}
W_{m(1)} = - \bigr(
\bar{Q}_i R^i_\alpha(A + Q) - \epsilon^{ab} E_{\alpha ab}^* \bigr) 
B^\alpha - \bigr( 
Q_{i a}^* R^i_\alpha(A + Q) + H_{\alpha a} \bigr) C^{\alpha a}.
\end{equation}

In order to explain how by iteration the $(n+1)$th order is obtained
from the $n$th order approximation let us introduce 
\begin{equation}
\label{A.3}
W_{m[n]} \equiv S_{\rm cl}(A + Q) + \sum_{k = 1}^n W_{m(k)}
\end{equation}
satisfying the $n$th order classical master equations
\begin{equation*}
\hbox{$\frac{1}{2}$} ( W_{m[n]}, W_{m[n]} )^a_{(k)} - 
V_m^a W_{m(k)} = 0,
\qquad
\hbox{$\frac{1}{2}$} \{ W_{m[n]}, W_{m[n]} \}_{A (k)} - 
V_A W_{m(k)} = 0,
\end{equation*}
for $k =1,2, \ldots, n$. Here, $( ~,~ )^a_{(k)}$ and $\{ ~,~ \}_{A(k)}$
denote the brackets of $k$th order in powers of the fields $B^\alpha$ and 
$C^{\alpha a}$. Furthermore, functionals $Y_{m(n + 1)}^a$ and $Y_{A(n + 1)}$ 
are constructed from $W_{m(k)}$, $k \leq n$, by the formulas
\begin{equation}
\label{A.4}
Y_{m(n + 1)}^a = - \hbox{$\frac{1}{2}$} ( W_{m[n]}, W_{m[n]} )_{(n + 1)}^a,
\qquad
Y_{A(n + 1)} = - \hbox{$\frac{1}{2}$} \{ W_{m[n]}, W_{m[n]} \}_{A(n + 1)}.
\end{equation}
Then, the $(n + 1)$th approximation $W_{m(n + 1)}$ of the solution (\ref{A.1})
is obtained by the help of the equations
\begin{equation}
\label{A.5}
\mathbf{O}_m^a W_{m(n + 1)} = Y_{m(n + 1)}^a,
\qquad
\mathbf{O}_A W_{m(n + 1)} = Y_{A(n + 1)}, 
\qquad
n \geq 1.
\end{equation}
where the operators $\mathbf{O}_m^a$ and $\mathbf{O}_A$ are given by 
\begin{align}
\label{A.6}
\mathbf{O}_m^a &= S_{\rm cl}(A + Q)_{,i} 
\frac{\delta}{\delta Q_{i a}^*} - \bigr(
Q_{i a}^* R^i_\alpha(A + Q) + H_{\alpha a} \bigr)
\frac{\delta}{\delta C_{\alpha ab}^*}  -
\epsilon^{ab} B^\alpha \frac{\delta_L}{\delta C^{\alpha b}}
\nonumber\\
& \quad - \bigr(
\bar{Q}_i R^i_\alpha(A + Q) - \epsilon^{bc} E_{\alpha bc}^* \bigr)
\frac{\delta}{\delta B_{\alpha a}^*} +
m^2 C^{\alpha a} \frac{\delta_L}{\delta B^\alpha} - V_m^a
\\
\label{A.7}
\mathbf{O}_A &= - (\sigma_A)_a^{~b} 
H_{\alpha b} \frac{\delta}{\delta F_{\alpha a}} - 
(\sigma_A)_a^{~b} 
C^{\alpha a} \frac{\delta_L}{\delta C^{\alpha b}} - V_A.
\end{align}
By construction they obey the $osp(1,2)$ superalgebra
\begin{equation*}
\{ \mathbf{O}_m^a, \mathbf{O}_m^b \} = m^2 (\sigma^A)^{ab} \mathbf{O}_A,
\qquad
[ \mathbf{O}_A, \mathbf{O}_m^a ] = - \mathbf{O}_m^b (\sigma_A)_b^{~a},
\qquad
[ \mathbf{O}_A, \mathbf{O}_B ] = - \epsilon_{AB}^{~~~~\!C} \mathbf{O}_C.
\end{equation*}
The functionals $Y_{m(n + 1)}^a$ and $Y_{A(n + 1)}$
obey the equations
\begin{align*}
\mathbf{O}_A Y_{B(n + 1)} - \mathbf{O}_B Y_{A(n + 1)} &= 
- \epsilon_{AB}^{~~~~\!C} Y_{C(n + 1)},
\\
\mathbf{O}_A Y_{m(n + 1)}^a - \mathbf{O}_m^a Y_{A(n + 1)} &=
- Y_{m(n + 1)}^b (\sigma_A)_b^{~a},
\\
\mathbf{O}_m^a Y_{m(n + 1)}^b + \mathbf{O}_m^b Y_{m(n + 1)}^a &=
m^2 (\sigma^A)^{ab} Y_{A(n + 1)},
\end{align*}
being the compability conditions for the Eqs. (\ref{A.5}). Note, that in 
solving Eqs. (\ref{A.5}) the $(n + 1)$th approximation $W_{m(n + 1)}$ is 
uniquely determined up to terms of the form 
$( \hbox{$\frac{1}{2}$} \epsilon_{ab} \mathbf{O}_m^b \mathbf{O}_m^a + m^2 )
X_{m(n + 1)}$, with $\mathbf{O}_A X_{m(n + 1)} = 0$, which still could be 
added to $W_{m(n + 1)}$; this is a consequence of the relations 
$\mathbf{O}_m^c ( 
\hbox{$\frac{1}{2}$} \epsilon_{ab} \mathbf{O}_m^b \mathbf{O}_m^a + m^2 ) =
\hbox{$\frac{1}{2}$} m^2 (\sigma^A)^c_{~d} \mathbf{O}_m^d \mathbf{O}_A$ and
$[ \mathbf{O}_A, 
\hbox{$\frac{1}{2}$} \epsilon_{ab} \mathbf{O}_m^b \mathbf{O}_m^a + m^2 ] = 0$.
But, in the present case it turns out that $X_{m(n + 1)} = 0$ for $n \geq 1$.

In order to obtain the higher order approximations, $W_{m(n + 1)}$, 
$n \geq 1$, one first has to construct the functionals $Y_{m(n + 1)}^a$ and 
$Y_{A(n + 1)}$ according to the rules (\ref{A.4}) which requires, by virtue of 
(\ref{A.3}), the explicit knowledge of all lower order approximations 
$W_{m(k)}$, $k = 1,2, \ldots, n$. After that, one has to determine the 
solutions of the equations (\ref{A.5}). Thereby, one has to employ all gauge 
structure relations, i.e., the algebra of the generators,
$R^i_{\alpha, j} R^j_\beta(A + Q) - R^i_{\beta, j} R^j_\alpha(A + Q) =
- R^i_\gamma(A + Q) f^\gamma_{\alpha\beta}$, the Jacobi identity 
$f^\delta_{\eta\alpha} f^\eta_{\beta\gamma} +
\hbox{cyclic perm.}(\alpha, \beta, \gamma) = 0$, and the antisymmetry 
of the structure constants $f^\gamma_{\alpha\beta}= - f^\gamma_{\beta\alpha}$. 

Omitting any details of the cumbersome algebraic manipulations, one gets 
at second order
\begin{align*}
Y_{m(2)}^a &= \hbox{$\frac{1}{2}$} Q_{i b}^* R^i_\alpha(A + Q)
f^\alpha_{\beta\gamma} C^{\beta a} C^{\gamma b} -
\hbox{$\frac{1}{2}$} \epsilon^{ab} Q_{i b}^*
R^i_{\alpha, j} R^j_\beta(A + Q) \epsilon_{cd} C^{\alpha c} C^{\beta d}
\\
& \quad + \hbox{$\frac{1}{2}$} \bar{Q}_i R^i_\alpha(A + Q)
f^\alpha_{\beta\gamma} B^\beta C^{\gamma a} -
\hbox{$\frac{1}{2}$} \bar{Q}_i \bigr( 
R^i_{\alpha, j} R^j_\beta(A + Q) +
R^i_{\beta, j} R^j_\alpha(A + Q) \bigr) B^\alpha C^{\beta a}, \\
Y_{A(2)} &= 0,
\end{align*}
which leads to
\begin{equation}
\label{A.8}
W_{m(2)} = - \hbox{$\frac{1}{2}$} E_{\gamma ab}^* f^\gamma_{\alpha\beta}
C^{\alpha a} C^{\beta b} +
\bar{E}_{\gamma a} f^\gamma_{\alpha\beta} B^\alpha C^{\beta a} + 
\hbox{$\frac{1}{2}$} \epsilon_{ab} \bar{Q}_i 
R^i_{\alpha, j} R^j_\beta(A + Q) C^{\beta b} C^{\alpha a}.
\end{equation}
Iterating the preceeding step at the next order one gets
\begin{align*}
Y_{m(3)}^a &= \hbox{$\frac{1}{4}$} \bar{Q}_i \bigr(
R^i_{\alpha, j} R^j_\beta(A + Q) + R^i_{\beta, j} R^j_\alpha(A + Q) \bigr)
f^\alpha_{\gamma\delta} \epsilon_{cd} 
C^{\delta c} C^{\beta d} C^{\gamma a}
\\
& \quad + \hbox{$\frac{1}{2}$} R^i_{\alpha, j} R^j_{\beta, k} R^k_\gamma(A + Q)
\epsilon_{cd} C^{\alpha c} C^{\beta d} C^{\gamma a} - 
\hbox{$\frac{1}{12}$} \epsilon^{ab} 
( C_{\alpha be}^* + C_{\alpha eb}^* ) 
f^\alpha_{\eta\beta} f^\eta_{\gamma\delta} \epsilon_{cd}
C^{\delta c} C^{\beta d} C^{\gamma e}
\\
& \quad + \hbox{$\frac{1}{4}$} \bar{E}_{\alpha b} 
( f^\alpha_{\eta\beta} f^\eta_{\gamma\delta} -
f^\alpha_{\eta\delta} f^\eta_{\gamma\beta} ) C^{\delta a} C^{\beta b}
B^\gamma + \hbox{$\frac{1}{4}$} \epsilon^{ab} \bar{E}_{\alpha b} 
f^\alpha_{\eta\beta} f^\eta_{\gamma\delta}
\epsilon_{cd} C^{\delta c} C^{\beta d} B^\gamma\\
Y_{A(3)} &= 0,
\end{align*}
which yields 
\begin{equation}
\label{A.9}
W_{m(3)} = \hbox{$\frac{1}{6}$} \epsilon_{cd} 
\bar{E}_{\alpha a} f^\alpha_{\eta\beta} f^\eta_{\gamma\delta} 
C^{\gamma a} C^{\delta c} C^{\beta d},
\end{equation}
for the third order approximation. The contributions to all the higher
approximations $W_{m(n)}$, $n \geq 4$, vanish identically due to the
gauge structure relations. Inserting into (\ref{A.1}) for $W_{m(n)}$
the explicit expressions (\ref{A.2}), (\ref{A.8}) and (\ref{A.9})  
we recover the solution (\ref{4.11}). Moreover, as a consequence of the 
boundary conditions, which means that the expansion (\ref{A.1}) must start 
with the classical action $S_{\rm cl}(A + Q)$, the solution (\ref{A.1}) is 
also type--I, type--II and type--III invariant.

Finally, let us justify the ansatz (\ref{3.6}). It has been established
in Ref.~\cite{3} 
for $Sp(2)$--symmetric irreducible theories that the characteristic
arbitrariness of a solution of the first master equation (\ref{A0}) for
$m = 0$ of a fixed order in the fields $C^{\alpha a}$ and $B^\alpha$ has the
form:
\begin{equation*}
\delta W^{(0)} = \hbox{$\frac{1}{2}$} \epsilon_{ab}
\mathbf{O}^b \mathbf{O}^a X,
\end{equation*}
where the nilpotent operators $\mathbf{O}^a$ are given by the 
$m$--independent part of (\ref{A.6}), while $X$ is an arbitrary
$Sp(2)$--scalar. This proof can be extended without any problems to the
$osp(1,2)$--symmetric case. The characteristic arbitrariness
of the solutions of both the master equations (\ref{A0}) for
$m \neq 0$ of a fixed order in the fields $C^{\alpha a}$ and $B^\alpha$ 
has the following form:
\begin{equation}
\label{A.10}
\delta W_m^{(0)} = ( \hbox{$\frac{1}{2}$} \epsilon_{ab}
\mathbf{O}_m^b \mathbf{O}_m^a + m^2 ) X
\quad 
{\rm with}
\quad
\mathbf{O}_A X = 0.
\end{equation}
If we consider now the variation $\partial S_m(\zeta)/ \partial \zeta$ 
of the functional ${\rm exp}\{(i/\hbar) S_m(\zeta) \}$ with $S_m(\zeta)$
satisfying (\ref{3.2}) it is obvious that the ansatz (\ref{3.6}) 
is just the counterpart of Eq.~(\ref{A.10}).


\section{Construction of the solutions (\ref{5.2}) and (\ref{5.3}) of the 
classical master equations}
\renewcommand{\theequation}{\thesection.\arabic{equation}}
\setcounter{equation}{0} 


In this Appendix we show how the solutions $S_m^{(0)}$ and 
$S_{m, {\rm ext}}^{(0)}$, Eqs.~(\ref{5.2}) and (\ref{5.3}), are 
constructed from the solutions $W_m^{(0)}$ and $W_{m, {\rm ext}}^{(0)}$, 
Eqs.~(\ref{4.11}) and (\ref{4.13}), by means of the generalized canonical 
transformation (\ref{3.16}). Thereby it will be proven that 
(\ref{5.3}) is the most general (gauge fixed) solution of the classical 
master equations with arbitrary dependence on the antifields.

According to (\ref{3.16}) and the commutative diagram (\ref{3.19}) the
solution $S_m(1) \equiv S_m^{(0)}$ is obtained from $S_m(0) \equiv
W_m^{(0)}$ by the expansion
\begin{align}
\label{B.1}
S_m^{(0)} &= \sum_{n = 0}^\infty \overset{(n)}{S}_{\!\!m},
\qquad
\overset{(0)}{S}_{\!\!m} \equiv W_m^{(0)},
\\
(n + 1) \overset{(n + 1)}{S}_{\!\!\!\!\!\!m} &= 
\hbox{\large $\frac{1}{2}$} \epsilon_{ab} \Bigr\{
\sum_{k = 0}^n ( \overset{(k)}{S}_{\!\!m}, 
( \overset{(n - k)}{S}_{\!\!\!\!\!\!m}, G )^a )^b -
( \overset{(n)}{S}_{\!\!m}, V_m^a G )^b 
\nonumber\\
& \quad~ -  V_m^b ( \overset{(n)}{S}_{\!\!m}, G )^a +
\delta_{n, 0} V_m^b V_m^a G \Bigr\} +
\delta_{n, 0} m^2 G,
\qquad
n \geq 0,
\nonumber
\end{align}
with 
\begin{equation*}
G = \hbox{$\frac{1}{2}$} ( 
\tilde{\sigma} g^{ij} \bar{Q}_i \bar{Q}_j + \tilde{\rho} \xi^{-1}
g^{\alpha\beta} \epsilon^{ab} \bar{E}_{\alpha a} \bar{E}_{\beta b} ),
\end{equation*}
where the antibrackets $( ~,~ )^a$ and the operators $V_m^a$ are defined 
in (\ref{2.23}) and (\ref{2.20}).

The evaluation of the second term in the expansion (\ref{B.1}) yields
\begin{align}
\label{B.2}
\overset{(1)}{S}_{\!\!m} = \, & \Bigr\{ 
\tilde{\sigma} g^{ij} \bar{Q}_j \frac{\delta}{\delta Q^i} + 
\tilde{\rho} \xi^{-1} g^{\alpha\beta} \epsilon^{ab}  \bigr( 
\bar{E}_{\beta b} \frac{\delta}{\delta C^{\alpha a}} + 
\hbox{\large $\frac{1}{2}$} E_{\beta ab}^* \frac{\delta}{\delta B^\alpha} 
\bigr) \Bigr\} W_m^{(0)} 
\\
& + \tilde{\sigma} X - 
\tilde{\rho} \xi^{-1} ( Y + g^{\alpha\beta} \epsilon^{ab} 
H_{\alpha a} \bar{E}_{\beta b} ) +
\hbox{\large $\frac{1}{4}$} \tilde{\rho} \xi^{-1} g^{\alpha\beta} 
\frac{\delta W_m^{(0)}}{\delta B^\alpha}
\frac{\delta W_m^{(0)}}{\delta B^\beta},
\nonumber
\end{align}
where the quantities $X$ and $Y$ are given by
\begin{equation*}
X \equiv \hbox{$\frac{1}{2}$} g^{ij} (
\epsilon^{ab} Q_{i a}^* Q_{j b}^* - m^2 \bar{Q}_i \bar{Q}_j ),
\qquad
Y \equiv g^{\alpha\beta} \epsilon^{ab} (
\hbox{$\frac{1}{2}$} \epsilon^{cd} E_{\alpha ac}^* E_{\beta bd}^* +
m^2 \bar{E}_{\alpha a} \bar{E}_{\beta d} ).
\end{equation*}
For the other terms, after tedious but straightforward calculations,
one obtains
\begin{align}
\label{B.3}
\overset{(2)}{S}_{\!\!m} = \,& \frac{1}{2!} \Bigr\{ 
\tilde{\sigma} g^{ij} \bar{Q}_j \frac{\delta}{\delta Q^i} + 
\tilde{\rho} \xi^{-1} g^{\alpha\beta} \epsilon^{ab}  \bigr( 
\bar{E}_{\beta b} \frac{\delta}{\delta C^{\alpha a}} +
\hbox{\large $\frac{1}{2}$} E_{\beta ab}^* \frac{\delta}{\delta B^\alpha} 
\bigr) \Bigr\}^2 W_m^{(0)} 
\\
& + \hbox{\large $\frac{1}{2}$} \tilde{\rho} \xi^{-1} g^{\alpha\beta} 
\frac{\delta W_m^{(0)}}{\delta B^\alpha} (
\tilde{\sigma} X_\beta + \tilde{\rho} \xi^{-1} Y_\beta ), 
\nonumber
\\
\label{B.4}
\overset{(3)}{S}_{\!\!m} = \,& \frac{1}{3!} \Bigr\{ 
\tilde{\sigma} g^{ij} \bar{Q}_j \frac{\delta}{\delta Q^i} + 
\tilde{\rho} \xi^{-1} g^{\alpha\beta} \epsilon^{ab} \bigr(  
\bar{E}_{\beta b} \frac{\delta}{\delta C^{\alpha a}} +
\hbox{\large $\frac{1}{2}$} E_{\beta ab}^* \frac{\delta}{\delta B^\alpha} 
\bigr) \Bigr\}^3 W_m^{(0)} 
\\
& + \hbox{\large $\frac{1}{4}$} \tilde{\rho} \xi^{-1} g^{\alpha\beta} 
( \tilde{\sigma} X_\alpha + \tilde{\rho} \xi^{-1} Y_\alpha )
( \tilde{\sigma} X_\beta + \tilde{\rho} \xi^{-1} Y_\beta ) 
\nonumber
\end{align}
and
\begin{align}
\label{B.5}
\overset{(n)}{S}_{\!\!m} = \,& \frac{1}{n!} \Bigr\{ 
\tilde{\sigma} g^{ij} \bar{Q}_j \frac{\delta}{\delta Q^i} + 
\tilde{\rho} \xi^{-1} g^{\alpha\beta} \epsilon^{ab} \bigr(  
\bar{E}_{\beta b} \frac{\delta}{\delta C^{\alpha a}} +
\hbox{\large $\frac{1}{2}$} E_{\beta ab}^* \frac{\delta}{\delta B^\alpha} 
\bigr) \Bigr\}^n W_m^{(0)}, 
\end{align}
for $n \geq 4$, with the abbreviations
\begin{equation}
\label{B.6}
X_\alpha \equiv - R_{\alpha, j}^i \bar{Q}_i g^{jk} \bar{Q}_k,
\qquad
Y_\alpha \equiv f^\beta_{\alpha\gamma}  
\epsilon^{ab} \bar{E}_{\beta a} g^{\gamma\delta} \bar{E}_{\delta b}.   
\end{equation}
Inserting into (\ref{B.1}) for $\overset{(n)}{S}_{\!\!m}$, $n \geq 1$, the
quantities (\ref{B.2}) -- (\ref{B.5}) it is easily seen 
that the resulting expression for $S_m^{(0)}$ can be cast into the form 
(\ref{5.2}). Notice, that adding to 
$W_m^{(0)}(A^i| Q^i, C^{\alpha a}, B^\alpha)$ only the first terms of the 
quantities (\ref{B.2}) -- (\ref{B.5}) gives the same 
functional $W_m^{(0)}(A^i| \tilde{Q}^i, \tilde{C}^{\alpha a}, 
\tilde{B}^\alpha)$ but with $Q^i$, $C^{\alpha a}$ and $B^\alpha$ being 
replaced according to the relations (\ref{5.1}).

The simplest way to construct the general solution $S_{m, {\rm ext}}^{(0)}$ 
consists in using the same generalized canonical transformation as in 
(\ref{B.1}) and 
merely changing the boundary conditions. Indeed, according to (\ref{3.16}) and
(\ref{3.19}) the solution $S_m(1) \equiv S_{m, {\rm ext}}^{(0)}$ is obtained 
from $S_m(0) \equiv W_{m, {\rm ext}}^{(0)}$ by the expansion
\begin{align}
\label{B.7}
S_{m, {\rm ext}}^{(0)} &= \sum_{n = 0}^\infty \overset{(n)}{S}_{\!\!m},
\qquad
\overset{(0)}{S}_{\!\!m} \equiv W_{m, {\rm ext}}^{(0)},
\\
(n + 1) \overset{(n + 1)}{S}_{\!\!\!\!\!\!m} &= 
\hbox{\large $\frac{1}{2}$} \epsilon_{ab} \Bigr\{
\sum_{k = 0}^n ( \overset{(k)}{S}_{\!\!m}, 
( \overset{(n - k)}{S}_{\!\!\!\!\!\!m}, G )^a )^b -
( \overset{(n)}{S}_{\!\!m}, V_m^a G )^b 
\nonumber\\
& \quad~ -  V_m^b ( \overset{(n)}{S}_{\!\!m}, G )^a +
\delta_{n, 0} V_m^b V_m^a G \Bigr\} +
\delta_{n, 0} m^2 G,
\qquad
n \geq 0.
\nonumber
\end{align}
For the second term in that expansion one gets 
\begin{align}
\label{B.8}
\overset{(1)}{S}_{\!\!m} = \, & \Bigr\{ 
\tilde{\sigma} g^{ij} \bar{Q}_j \frac{\delta}{\delta Q^i} + 
\tilde{\rho} \xi^{-1} g^{\alpha\beta} \epsilon^{ab}  \bigr( 
\bar{E}_{\beta b} \frac{\delta}{\delta C^{\alpha a}} + 
\hbox{\large $\frac{1}{2}$} E_{\beta ab}^* \frac{\delta}{\delta B^\alpha} 
\bigr) \Bigr\} W_{m, {\rm ext}}^{(0)} 
\\
& + \tilde{\sigma} X - 
\tilde{\rho} \xi^{-1} ( Y + g^{\alpha\beta} \epsilon^{ab} 
H_{\alpha a} \bar{E}_{\beta b} ) +
\hbox{\large $\frac{1}{4}$} \tilde{\rho} \xi^{-1} g^{\alpha\beta} 
\frac{\delta W_{m, {\rm ext}}^{(0)}}{\delta B^\alpha}
\frac{\delta W_{m, {\rm ext}}^{(0)}}{\delta B^\beta}
\nonumber
\\
\intertext{with the same expression for $X$ and $Y$ as in (\ref{B.2}). 
For the other terms one obtains}
\label{B.9}
\overset{(2)}{S}_{\!\!m} = \,& \frac{1}{2!} \Bigr\{ 
\tilde{\sigma} g^{ij} \bar{Q}_j \frac{\delta}{\delta Q^i} + 
\tilde{\rho} \xi^{-1} g^{\alpha\beta} \epsilon^{ab}  \bigr( 
\bar{E}_{\beta b} \frac{\delta}{\delta C^{\alpha a}} +
\hbox{$\frac{1}{2}$} E_{\beta ab}^* \frac{\delta}{\delta B^\alpha} \bigr) 
\Bigr\}^2 W_{m, {\rm ext}}^{(0)} 
\\
& + \hbox{\large $\frac{1}{4}$} \tilde{\rho} \xi^{-1} g^{\alpha\beta} \Bigr\{
\tilde{\rho} \frac{\delta W_{m, {\rm ext}}^{(0)}}{\delta B^\alpha}
\frac{\delta W_{m, {\rm ext}}^{(0)}}{\delta B^\beta} +
2 \frac{\delta W_{m, {\rm ext}}^{(0)}}{\delta B^\alpha} (
\tilde{\sigma} \tilde{X}_\beta + \tilde{\rho} \xi^{-1} \tilde{Y}_\beta ) 
\Bigr\},
\nonumber
\\
\intertext{and}
\label{B.10}
\overset{(n)}{S}_{\!\!m} = \,& \frac{1}{n!} \Bigr\{ 
\tilde{\sigma} g^{ij} \bar{Q}_j \frac{\delta}{\delta Q^i} + 
\tilde{\rho} \xi^{-1} g^{\alpha\beta} \epsilon^{ab} \bigr(  
\bar{E}_{\beta b} \frac{\delta}{\delta C^{\alpha a}} +
\hbox{$\frac{1}{2}$} E_{\beta ab}^* \frac{\delta}{\delta B^\alpha} \bigr)
\Bigr\}^n W_{m, {\rm ext}}^{(0)} 
\\
& + \hbox{\large $\frac{1}{4}$} \tilde{\rho}^{n - 2} 
\xi^{-1} g^{\alpha\beta} \Bigr\{
\tilde{\rho}^2 \frac{\delta W_{m, {\rm ext}}^{(0)}}{\delta B^\alpha}
\frac{\delta W_{m, {\rm ext}}^{(0)}}{\delta B^\beta} +
2 \tilde{\rho} \frac{\delta W_{m, {\rm ext}}^{(0)}}{\delta B^\alpha} (
\tilde{\sigma} \tilde{X}_\beta + \tilde{\rho} \xi^{-1} \tilde{Y}_\beta )
\nonumber
\\
&\qquad\qquad\qquad\qquad\!\!\! + 
( \tilde{\sigma} \tilde{X}_\alpha + 
\tilde{\rho} \xi^{-1} \tilde{Y}_\alpha )
( \tilde{\sigma} \tilde{X}_\beta + 
\tilde{\rho} \xi^{-1} \tilde{Y}_\beta ) \Bigr\},
\nonumber
\end{align}
for $n \geq 3$, where $\tilde{X}_\alpha$ and $\tilde{Y}_\alpha$ are obtained 
from the expressions $X_\alpha$ and $Y_\alpha$, Eqs.~(\ref{B.6}), according
to
\begin{equation*}
\tilde{X}_\alpha \equiv X_\alpha + R_\alpha^i(A) \bar{Q}_i,
\qquad
\tilde{Y}_\alpha \equiv Y_\alpha + \xi \epsilon^{ab} E_{\alpha ab}^*.
\end{equation*}
Inserting into (\ref{B.7}) for $\overset{(n)}{S}_{\!\!m}$, $n \geq 1$,
the quantities (\ref{B.8}) -- (\ref{B.10}) and using the equality 
$(1 - \tilde{\rho})^{-1} = \sum_{n = 0}^\infty \tilde{\rho}^n$ one simply 
estabishes that the resulting expression for $S_{m, {\rm ext}}^{(0)}$ can
be expressed in the form (\ref{5.3}).


\end{appendix}

\end{document}